\documentclass[a4paper, 11pt]{article}

\usepackage{authblk}

\usepackage{graphicx}
\usepackage{subfig} 

\usepackage{amsmath}
\usepackage{amsfonts}
\usepackage{amssymb}
\usepackage{amsthm}
\usepackage{amsopn}
\usepackage[mathscr]{eucal}
\usepackage{array}
\usepackage{mathtools}
\usepackage{cancel}

\usepackage{tabularx}

\usepackage{setspace}

\usepackage[numbers,square]{natbib}

\usepackage{url}

\usepackage[siunitx]{circuitikz}
\usepackage{tikz}
\usetikzlibrary{arrows}

\usepackage{pgfplots}
\pgfplotsset{compat=1.13}

\usepackage[a4paper, top=1in, bottom=1in, left=1in, right=1in ]{geometry}

\usepackage{framed}
\makeatletter
\def\th@newremark{\th@remark\thm@headfont{\bfseries}}
\makeatletter

\theoremstyle{newremark}
\newtheorem{remark}{Remark}[section]

\renewcommand{\complement}{\mathsf{c}}

\newcommand{\gradft}{\nabla_{t}\mathbf{f}}
\newcommand{\btrip}{(\mathbf{x} , \boldsymbol{\theta} ,t )}
\newcommand{\bkp}{\pmb{\boldsymbol{\varkappa}}}
\newcommand{\bxi}{\boldsymbol{\Xi}}

\newcommand{\T}{\mathsf{T}}
\newcommand{\eye}{\mathbf{I}}
\newcommand{\zero}{\mathbf{O}}
\DeclareMathOperator{\cov}{\mathbf{Cov}}
\DeclareMathOperator{\mean}{\mathbf{Mean}}

\newcommand{\boldy}{\mathbf{y}}
\newcommand{\boldx}{\mathbf{x}}
\newcommand{\boldS}{\mathbf{S}}

\newcommand{\btheta}{\boldsymbol{\theta}}
\newcommand{\bthetas}{\boldsymbol{\theta}^{\{\mathcal{S}\}}}
\newcommand{\bthetasc}{\boldsymbol{\theta}^{\{\mathcal{S}^\complement\}}}

\newcommand{\bthetaw}{\boldsymbol{\theta}^{\{\mathcal{W}\}}}
\newcommand{\bthetaswc}{\boldsymbol{\theta}^{\{(\mathcal{S}\cup\mathcal{W})^\complement\}}}
\newcommand{\suwc}{{(\mathcal{S}\cup\mathcal{W})^\complement}}

\newcommand{\ccn}{\boldsymbol{\mathcal{C}}_n}
\newcommand{\ccns}{\boldsymbol{\mathcal{C}}_n^{\{\mathcal{S}\}}}
\newcommand{\ccnsgw}{\boldsymbol{\mathcal{C}}_n^{\{\mathcal{S|W}\}}}

\newcommand{\sss}{\mathbf{S}^{\{\mathcal{S}\}}}
\newcommand{\ssst}{\mathbf{S}^{\{\mathcal{S}\}^\T}}

\newcommand{\sssc}{\mathbf{S}^{\{\mathcal{S}^\complement\}}}
\newcommand{\sssct}{\mathbf{S}^{\{\mathcal{S}^\complement\}^\T}}

\newcommand{\sssuwc}{\mathbf{S}^{\{(\mathcal{S}\cup\mathcal{W})^\complement\}}}
\newcommand{\sssuwct}{\mathbf{S}^{\{(\mathcal{S}\cup\mathcal{W})^\complement\}^\T}}

\newcommand{\thetass}{\boldsymbol\theta^{\{\mathcal{S}\}}}
\newcommand{\thetassc}{\boldsymbol\theta^{\{\mathcal{S}^\complement\}}}

\newcommand{\delthetass}{\nabla_{{\boldsymbol{\theta}}^{\{\mathcal{S}\}}} \, \mathbf{x}}
\newcommand{\delthetassc}{\nabla_{{\boldsymbol{\theta}}^{\{\mathcal{S}^\complement\}}} \, \mathbf{x}}

\newcommand{\delthetassuwc}{\nabla_{{\boldsymbol{\theta}}^{\{(\mathcal{S}\cup\mathcal{W})^\complement\}}} \, \mathbf{x}}

\newcommand{\inns}{\mathcal{I}^{\{\mathcal{S}\}}_n}

\newcommand{\innsgw}{\mathcal{I}^{\{\mathcal{S|W}\}}_n}

\DeclareMathAlphabet{\mathsfit}{\encodingdefault}{\sfdefault}{m}{sl}
\newcommand{\cards}{\mathsfit{s}}
\newcommand{\cardw}{\mathsfit{w}}

\newcommand{\iext}{I_\mathrm{ext}}
\newcommand{\gna}{g_{Na}}
\newcommand{\gk}{g_{K}}
\newcommand{\gl}{g_{L}}

\title{Information sensitivity functions to assess parameter information gain and identifiability of dynamical systems}

\author[1]{Sanjay Pant}

\affil[1]{Zienkiewicz Centre for Computational Engineering, Swansea University, UK.}

\date{\today}

\begin{document}
\maketitle

\begin{abstract}
    \noindent
    A new class of functions, called the `Information sensitivity functions' (ISFs), which quantify the information gain about the parameters through the measurements/observables of a dynamical system are presented. These functions can be easily computed through classical sensitivity functions alone and are based on Bayesian and information-theoretic approaches. While marginal information gain is quantified by decrease in differential entropy, correlations between arbitrary sets of parameters are assessed through mutual information. For individual parameters these information gains are also presented as marginal posterior variances, and, to assess the effect of correlations, as conditional variances when other parameters are given. The easy to interpret ISFs can be used to a) identify time-intervals or regions in dynamical system behaviour where information about the parameters is concentrated; b) assess the effect of measurement noise on the information gain for the parameters; c) assess whether sufficient information in an experimental protocol (input, measurements, and their frequency) is available to identify the parameters; d) assess correlation in the posterior distribution of the parameters to identify the sets of parameters that are likely to be indistinguishable; and e) assess identifiability problems for particular sets of parameters.

\end{abstract}

\section{Introduction}
Sensitivity analysis \cite{saltelli2000sensitivity} has been widely used to determine how the parameters of a dynamical system influence its outputs. When one or more outputs are measured (observed), it quantifies the variation of the observations with respect to the parameters to determine which parameters are most and least influential towards the measurements. Therefore, when performing an inverse problem of estimating the parameters from the measurements, sensitivity analysis is widely used to fix the least influential parameters (as their effect on the measurements is insignificant and removing them reduces the dimensionality of the inverse problem) while focussing on estimation of the most influential parameters. 
Sensitivity analysis is also used to assess the question of parameter identifiability, \emph{i.e.} how easy or difficult is it to identify the parameters from the measurements. This is primarily based on the idea the if the observables are highly sensitive to perturbations in certain parameters then these parameters are likely to be identifiable, and if the observables are insensitive then the parameters are likely to be unidentifiable. However, the magnitude of the sensitivities is hard to interpret, except in the trivial case when the sensitivities are identically zero. Lastly, parameter identifiability based on sensitivity analysis also assesses correlation/dependence between the parameters---through  principle component analysis \cite{degenring2004sensitivity}, correlation method \cite{jacquez1985compartmental}, orthogonal method \cite{yao2003modeling}, and the eigenvalue method \cite{quaiser2009systematic}---to identify which pairs of parameters, owing to the high correlation, are likely to be indistinguishable from each other (also see \cite{miao2011identifiability} and the referenced therein). Another method to assess correlations is based on the Fisher Information Matrix \cite{frieden2004science,miao2011identifiability,rodriguez2006novel}, which can be derived from asymptotic analysis of non-linear least squares estimators \cite{seber1989nonlinear,banks2009inverse}. 
Thomaseth and Cobeli extended the classical sensitivity functions to `generalized sensitivity functions' (GSFs) which assess information gain about the parameters from the measurements. This method has been widely used to assess identifiability of dynamical systems \cite{banks2009inverse,bai2007stochastic,batzel2009modelling,banks2010generalized,pant2014methodological}, where regions of high information gain show a sharp increase in the GSFs while oscillations imply correlation with other parameters. There are two drawbacks of GSFs: first, that they are designed to start at 0 and end at 1, which leads to the so called `force-to-one' phenomenon, where even in the absence of information about the parameters the GSFs are forces to end at a value of 1; and second, oscillations in GSFs can be hard to interpret in terms of identifying which sets of parameters are correlated. Based on a pure information-theoretic approach Pant and Lombardi \cite{pant2015Information} proposed to compute information gain through a decrease in Shannon entropy, which alleviated the shortcomings of GSFs. However, since their method relies on a Monte Carlo type method the computational effort associated with the computation of information gains can be quite large. 
In this article, a novel method which combines the method of Pant and Lombardi \cite{pant2015Information} with the classical sensitivity functions to compute information gain about the parameters is presented. The new functions are collectively called `Information sensitivity functions' (ISFs), which assess parameter information gain through sensitivity functions alone, thereby eliminating the need for Monte Carlo runs. These functions (i) are based on Bayesian/information-theoretic methods and do not rely on asymptotic analysis; (ii) are monotonically non-decreasing and therefore do not oscillate; (iii) can assess regions of high information content for individual parameters; iv) can assess parameter correlations between an arbitrary set of parameters; (v) can reveal potential problems in identifiability of system parameters; (vi) can assess the effect of experimental protocol on the inverse problem, for example, which outputs are measured, associated measurement noise, and measurement frequency; and (vii) are easily interpretable. 

In what follows, first the theoretic developments are presented in sections \ref{sec_dynamical_system_sensitivity}--\ref{sec_inforamtion_gain}, followed by their application to three different dynamical systems in section \ref{sec_results_and_discussion}. The three examples are chosen from different areas in mathematical biosciences: i) a Windkessel model, which is widely used a boundary condition in computational fluid dynamics simulations of haemodynamics; ii) the Hodgkin-Huxley model for a biological neuron, which has formed the basis for a variety of ionic models describing excitable tissues; and iii) a kinetics model for the Influenza A virus.

\section{The dynamical system and sensitivity equations}
\label{sec_dynamical_system_sensitivity}
Consider the following dynamical system governed by a set of parameterised ordinary differential equations (ODEs)
\begin{equation}
\label{eqn_forward}
    \dot{\mathbf{x}} = \mathbf{f} (\mathbf{x} , \boldsymbol{\theta} ,t ) \qquad  \text{ with } \qquad \mathbf{x} (t_0) = \mathbf{x}_0,
\end{equation}
where $t$ represents time, $\mathbf{x} \in \mathbb{R}^{d} $ is the state vector, $\boldsymbol{\theta} \in \mathbb{R}^{p} $ is the parameter vector, the function $\mathbf{f} : \mathbb{R}^{d+p+1} \to \mathbb{R}^{d} $ represents the dynamics, and $\mathbf{x}_0 $ represents the initial condition at time $t_0$. The initial conditions may depend on the parameters, and therefore
\begin{equation}
    \mathbf{x} (t_0) = \mathbf{x}_0(\btheta).
\end{equation}
The above representation subsumes the case where the initial condition may itself be seen as a parameter.
The RHS of the dynamical system, equation \eqref{eqn_forward}, can be linearised at at a reference point $(\mathbf{x}_r,\boldsymbol{\theta}_r,t_r)$, to obtain
\begin{equation}
    \label{eqn_discretised_forward_initial}
    \dot{\mathbf{x}} = \mathbf{f} (\mathbf{x} ,\boldsymbol{\theta}, t )\Bigr\rvert_{r}
    + \nabla_{\mathbf{x}} \mathbf{f}(\mathbf{x} ,\boldsymbol{\theta}, t )\Bigr\rvert_{r} (\mathbf{x} - \mathbf{x}_{r})  + \nabla_{\boldsymbol{\theta}} \mathbf{f}(\mathbf{x} ,\boldsymbol{\theta}, t )\Bigr\rvert_{r} (\boldsymbol{\theta} - \boldsymbol{\theta}_{r})
    + \gradft(\mathbf{x},\boldsymbol{\theta},t)\Bigr\rvert_{r} (t-t_r)
    + \mathcal{O}(2),
\end{equation}
where $(\cdot)\Bigr\rvert_r$ represents $(\cdot)$ evaluated at 
the reference point.
Henceforth, in order to be concise, the explicit dependence of $\mathbf{f}\btrip$ on its arguments is omitted and $\mathbf{f}$, without any arguments, is used to denote $\mathbf{f}\btrip$. Following this notation, equation \eqref{eqn_discretised_forward_initial} is concisely written as
\begin{equation}
    \label{eqn_discretised_forward_initial_simplified}
    \dot{\mathbf{x}} = \mathbf{f} \Bigr\rvert_{r}
    + \nabla_{\mathbf{x}} \mathbf{f}\Bigr\rvert_{r} (\mathbf{x} - \mathbf{x}_{r})  
    + \nabla_{\boldsymbol{\theta}} \mathbf{f}\Bigr\rvert_{r} (\boldsymbol{\theta} - \boldsymbol{\theta}_{r})
    + \gradft\Bigr\rvert_{r} (t-t_r)
    + \mathcal{O}(2).
\end{equation}
%
The above linearisation will be used in the next section to study the evolution of the state covariance matrix with time. 
Let $\mathbf{S} \in \mathbb{R}^{d\times p} $ denote the matrix of sensitivity functions for the system in equation \eqref{eqn_forward}, \emph{i.e.}  $\mathbf{S} = \nabla_{\boldsymbol{\theta}} \, \mathbf{x} $, or
\begin{equation}
S_{i,j} = \frac{\partial x_i}{\partial \theta_j}.
\end{equation}
It is well known that $\mathbf{S}$ satisfies the following ODE system, which can be obtained by applying the chain rule of differentiation to equation \eqref{eqn_forward}:
\begin{equation}
    \label{eqn_sensitivity}
    \dot{\mathbf{S}} = \Big( \nabla_{\mathbf{x}} \mathbf{f} \btrip \Big) \mathbf{S} + \nabla_{\boldsymbol{\theta}} \mathbf{f} \btrip
    \qquad \text{ with } \qquad
    \mathbf{S} (t_0) = \nabla_{\boldsymbol{\theta}} \left( \mathbf{x}_0(\boldsymbol{\theta} ) \right).
\end{equation}
 The goal is to relate the evolution of the sensitivity matrix to the evolution of the covariance of the joint vector of the state and the parameters.
Let the subscript $n$ denote all quantities at time $t_n$; for example, $\boldx_n$ denotes the state vector at time  $t_n$, $\mathbf{S}_n$ the corresponding sensitivity matrix, and so on. 
To relate the sensitivity matrix $\boldS_{n+1}$ at time $t_{n+1}$ with $\boldS_n$, a first order discretisation of equation \eqref{eqn_sensitivity} is considered
\begin{equation}
    \label{eqn_linearised_sensitivity}
    \frac{\mathbf{S}_{n+1}-\mathbf{S}_{n}  }{\Delta t} = \nabla_{\mathbf{x}} \mathbf{f}\Bigr\rvert_{n} \mathbf{S}_{n} + \nabla_{\boldsymbol{\theta}} \mathbf{f}\Bigr\rvert_{n},
\end{equation}
and, therefore, the matrix product $\mathbf{S}_{n+1} \mathbf{S}^{\T}_{n+1}$ can be written as
\begin{align}
    \label{eqn_SSt_recur}
    \mathbf{S}_{n+1} \mathbf{S}^{\T}_{n+1}   &= \phantom{+} \mathbf{S}_{n} \mathbf{S}^{\T}_{n} &+& \quad \mathbf{S}_{n} \mathbf{S}^{\T}_{n} \left( \nabla^\T_{\mathbf{x}} \mathbf{f}\Bigr\rvert_{n}  \right) \Delta t &+& \quad \mathbf{S}_{n} \left( \nabla^\T_{\boldsymbol{\theta}} \mathbf{f}\Bigr\rvert_{n}  \right) \Delta t \\
                                             \nonumber &\phantom{=} + \left( \nabla_{\mathbf{x}} \mathbf{f}\Bigr\rvert_{n}  \right) \mathbf{S}_{n} \mathbf{S}^{\T}_{n} \, \Delta t &+& \left( \nabla_{\mathbf{x}} \mathbf{f}\Bigr\rvert_{n}  \right) \mathbf{S}_{n}  \mathbf{S}^{\T}_{n} \left( \nabla^\T_{\mathbf{x}} \mathbf{f}\Bigr\rvert_{n}  \right) \Delta t^2 &+& \left( \nabla_{\mathbf{x}} \mathbf{f}\Bigr\rvert_{n}  \right) \mathbf{S}_{n} \left( \nabla^\T_{\boldsymbol{\theta}} \mathbf{f}\Bigr\rvert_{n}  \right)\Delta t^2\\
                                             \nonumber &\phantom{=}+ \left( \nabla_{\boldsymbol{\theta}} \mathbf{f}\Bigr\rvert_{n}  \right) \mathbf{S}^{\T}_{n} \, \Delta t &+& \left( \nabla_{\boldsymbol{\theta}} \mathbf{f}\Bigr\rvert_{n}  \right)\mathbf{S}^{\T}_{n} \left( \nabla^\T_{\mathbf{x}} \mathbf{f}\Bigr\rvert_{n}  \right) \Delta t^2 &+& \left( \nabla_{\boldsymbol{\theta}} \mathbf{f}\Bigr\rvert_{n}  \right) \left( \nabla^\T_{\boldsymbol{\theta}} \mathbf{f}\Bigr\rvert_{n}  \right) \Delta t^2.
\end{align}
 
Next, it is hypothesised that under certain conditions $\mathbf{S}_{n+1} \mathbf{S}^{\T}_{n+1}$ can be seen as the covariance matrix of the state vector at time $t_{n+1}$. These developments are presented in the next two sections.

\section{Forward propagation of uncertainty}
\label{sec:forward_propagation}
Since the objective is to study the relationship between the parameters and the state vector, a joint vector of all the state vectors until the current time $t_n$ and the parameter vector is considered.
Assume that at time $t_n$, this joint vector $[\mathbf{x}^{\T}_{n}, \mathbf{x}^{\T}_{n-1} , \ldots, \mathbf{x}^{\T}_{0}, \boldsymbol{\theta}^{\T} ]^{\T}$ is distributed according to a multivariate Normal distribution as follows
\begin{equation}
    \label{eqn_jd_x_theta_n}
\begin{bmatrix}
    \mathbf{x}_{n} \\
    \mathbf{x}_{n-1} \\
    \vdots \\
    \mathbf{x}_{0} \\
    \boldsymbol{\theta}
\end{bmatrix} \sim
\mathcal{N}
\left( \boldsymbol{\mu}_n =
    \begin{bmatrix}
        \boldsymbol{\mu}_{\mathbf{x}_{n}} \\[1ex]
        \boldsymbol{\mu}_{\mathbf{x}_{n-1}} \\[1ex]
        \vdots \\[1ex]
            \boldsymbol{\mu}_{\mathbf{x}_{0}} \\[1ex]
            \boldsymbol{\mu}_{\boldsymbol{\theta}}
        \end{bmatrix} ,\quad
 \boldsymbol{\Sigma}_n =
\begin{bmatrix}
    \boldsymbol{\Sigma}_{n,n}                         & \boldsymbol{\Sigma}_{n,n-1}      & \cdots & \boldsymbol{\Sigma}_{n,0}                         & \boldsymbol{\Lambda}_{n,\boldsymbol{\theta}} \\[1ex]
    \boldsymbol{\Sigma}_{n-1,n}                  & \boldsymbol{\Sigma}_{n-1,n-1}    & \cdots & \boldsymbol{\Sigma}_{n-1,0}                       & \boldsymbol{\Lambda}_{n-1,\boldsymbol{\theta}} \\[1ex]
    \vdots                                            & \vdots                           & \ddots & \vdots                                            & \vdots \\[1ex]
    \boldsymbol{\Sigma}_{0,n}                         & \boldsymbol{\Sigma}_{0,n-1} & \cdots & \boldsymbol{\Sigma}_{0,0}                         & \boldsymbol{\Lambda}_{0,\boldsymbol{\theta}} \\[1ex]
    \boldsymbol{\Lambda}_{\boldsymbol{\theta},n} & \boldsymbol{\Lambda}_{\boldsymbol{\theta},n-1} & \cdots & \boldsymbol{\Lambda}_{\boldsymbol{\theta},0} & \boldsymbol{\Sigma}_{\boldsymbol{\theta},\boldsymbol{\theta}}
\end{bmatrix}
\right).
\end{equation}

To obtain the joint distribution of $[\mathbf{x}^{\T}_{n+1}, \mathbf{x}^{\T}_{n} , \ldots, \mathbf{x}^{\T}_{0}, \boldsymbol{\theta}^{\T} ]^{\T}$ (all the state vectors until time $t_{n+1}$ and the parameater vector), the linearised dynamical system, equation \eqref{eqn_discretised_forward_initial_simplified}, is utilised. Considering the reference point $(\mathbf{x}_r,\boldsymbol{\theta}_r,t_r)$ in equation \eqref{eqn_discretised_forward_initial_simplified} to be $(\boldsymbol{\mu}_{\mathbf{x}_n},\boldsymbol{\mu}_{\boldsymbol{\theta}},t_n)$, \emph{i.e.} considering the linearisation around the mean values of the parameter vector and the state at time $t_n$, one obtains
\begin{equation}
    \label{eqn_discretised_forward_mean_app}
    \dot{\mathbf{x}} = \mathbf{f}\Bigr\rvert_{n}  
    + \nabla_{\mathbf{x}} \mathbf{f}\Bigr\rvert_{n} \left( \mathbf{x} - \boldsymbol{\mu}_{\mathbf{x}_{n}}  \right)   
    + \nabla_{\boldsymbol{\theta}} \mathbf{f}\Bigr\rvert_{n} \left(\boldsymbol{\theta} - \boldsymbol{\mu}_{\boldsymbol{\theta}} \right)
    + \gradft\Bigr\rvert_{n} (t-t_n) 
    + \mathcal{O}(2).
\end{equation}
Ignoring the higher order terms, and employing a forward Euler discretisation, one obtains 
\begin{equation}
    \label{eqn_discretised_forward_mean}
    \mathbf{x}_{n+1} \approx \mathbf{x}_{n} + \mathbf{f}\Bigr\rvert_{n} \Delta t + \nabla_{\mathbf{x}} \mathbf{f}\Bigr\rvert_{n} \left( \mathbf{x}_n - \boldsymbol{\mu}_{\mathbf{x}_{n}}  \right) \Delta t  + \nabla_{\boldsymbol{\theta}} \mathbf{f}\Bigr\rvert_{n} \left(\boldsymbol{\theta} - \boldsymbol{\mu}_{\boldsymbol{\theta}} \right) \Delta t.
\end{equation}
\begin{remark}
    $\mathbf{x}_{n+1}$ is completely determined by $\mathbf{x}_{n}$ and $\boldsymbol{\theta}$, \emph{i.e.} given $\mathbf{x}_{n}$ and $\boldsymbol{\theta}$ nothing more can be learned about $\mathbf{x}_{n+1}$. Hence, the forward propagation forms a Markov chain.
\end{remark}

\begin{remark}
    $\mathbf{f}\Bigr\rvert_{n}$, $\nabla_{\mathbf{x}} \mathbf{f}\Bigr\rvert_{n}$, $\nabla_{\boldsymbol{\theta}} \mathbf{f}\Bigr\rvert_{n}$ are evaluated at $(\boldsymbol{\mu}_{\mathbf{x}_n}, \boldsymbol{\mu}_{\boldsymbol{\theta}},t_n)$.
\end{remark}

\begin{remark}
    In equation \eqref{eqn_jd_x_theta_n}, $\boldsymbol{\Sigma}_{\alpha,\beta} = \boldsymbol{\Sigma}^{\T}_{\beta,\alpha}$ and  $\boldsymbol{\Lambda}_{\alpha,\beta} = \boldsymbol{\Lambda}^{\T}_{\beta,\alpha}$.
\end{remark}

The joint vector $[\mathbf{x}^{\T}_{n+1}, \mathbf{x}^{\T}_{n} , \ldots, \mathbf{x}^{\T}_{0}, \boldsymbol{\theta}^{\T} ]^{\T}$ can be written from equations \eqref{eqn_jd_x_theta_n} and \eqref{eqn_discretised_forward_mean} as
\begin{equation}
    \label{eqn_prop_state_theta}
\begin{bmatrix}
    \mathbf{x}_{n+1} \\[1ex]
    \mathbf{x}_{n} \\[1ex]
    \mathbf{x}_{n-1} \\[1ex]
    \vdots \\[1ex]
    \mathbf{x}_{0} \\[1ex]
    \boldsymbol{\theta}
\end{bmatrix} \approx
\underbrace{
\begin{bmatrix}
    \eye_{d} + \nabla_{\mathbf{x}} \mathbf{f}\Bigr\rvert_{n}  \Delta t & \zero_{d,d} & \zero_{d,d} & \cdots & \zero_{d,d} & \nabla_{\boldsymbol{\theta}} \mathbf{f}\Bigr\rvert_{n} \Delta t \\[1ex]
    \eye_{d}                                                           & \zero_{d,d} & \zero_{d,d} & \cdots & \zero_{d,d} & \zero_{d,p} \\[1ex]
    \zero_{d,d}                                                        & \eye_{d}    & \zero_{d,d} & \cdots & \zero_{d,d} & \zero_{d,p} \\[1ex]
    \vdots                                                             & \vdots      & \vdots      & \ddots & \vdots      & \vdots \\[1ex]
    \zero_{d,d}                                                        & \zero_{d,d} & \zero_{d,d} & \cdots & \eye_{d}    & \zero_{d,p}\\[1ex]
    \zero_{p,d}                                                        & \zero_{p,d} & \zero_{p,d} & \cdots & \zero_{p,d} & \eye_{p}
\end{bmatrix}
}_{\textstyle \mathbf{F}_{n}}
    \begin{bmatrix}
    \mathbf{x}_{n} \\[1ex]
    \mathbf{x}_{n-1} \\[1ex]
    \vdots \\[1ex]
    \mathbf{x}_{0} \\[1ex]
    \boldsymbol{\theta}
\end{bmatrix}
+
\underbrace{
\begin{bmatrix}
    \mathbf{C}_n\\[1ex]
        \zero_{d,1} \\[1ex]
        \zero_{d,1} \\[1ex]
        \vdots \\[1ex]
        \zero_{d,1} \\[1ex]
        \zero_{p,1} \\[1ex]
\end{bmatrix}
}_{\textstyle \mathbf{g}_{n}},
\end{equation}
where $\eye_q$ represents an Identity matrix of size $q$, $\zero_{q,r}$ represents a zero matrix of size $q\times r$, and 
\begin{equation}
\mathbf{C}_n = \mathbf{f}\Bigr\rvert_{n} \Delta t - \nabla_{\mathbf{x}} \mathbf{f}\Bigr\rvert_{n} \boldsymbol{\mu}_{\mathbf{x}_{n}} \; \Delta t  - \nabla_{\boldsymbol{\theta}} \mathbf{f}\Bigr\rvert_{n} \boldsymbol{\mu}_{\boldsymbol{\theta}} \; \Delta t 
\end{equation}
is a term that does not depend on $\mathbf{x}_{n}$ and $\boldsymbol{\theta}$. The distribution of $[\mathbf{x}^{\T}_{n+1}, \mathbf{x}^{\T}_{n} , \ldots, \mathbf{x}^{\T}_{0}, \boldsymbol{\theta}^\T ]^{\T}$ can be written from equation \eqref{eqn_prop_state_theta} as
\begin{equation}
    \label{eqn_jd_forward}
[\mathbf{x}^{\T}_{n+1}, \mathbf{x}^{\T}_{n} , \ldots, \mathbf{x}^{\T}_{0}, \boldsymbol{\theta} ]^{\T}
\sim \mathcal{N}
\left( \boldsymbol{\mu}_{n+1} = \mathbf{F}_n \boldsymbol{\mu}_n + \mathbf{g}_n
   , \quad
\boldsymbol{\Sigma}_{n+1} = \mathbf{F}_n \; \boldsymbol{\Sigma}_{n} \; \mathbf{F}^{\T}_n
\right),
\end{equation}
and the covariance $\boldsymbol{\Sigma}_{n+1}$ can be expanded as

\begin{equation}
\label{eqn_large_signp1}
\boldsymbol{\Sigma}_{n+1} =
\begin{bmatrix}
    \boldsymbol{\Sigma}_{n+1,n+1}                     & \boldsymbol{\Sigma}_{n+1,n}                       & \boldsymbol{\Sigma}_{n+1,n-1}                         & \cdots & \boldsymbol{\Sigma}_{n+1,0}                       & \boldsymbol{\Lambda}_{n+1,\boldsymbol{\theta}} \\[2ex]
    \boldsymbol{\Sigma}^{\T}_{n+1,n}                  & \boldsymbol{\Sigma}_{n,n}                         & \boldsymbol{\Sigma}_{n,n-1}                         & \cdots & \boldsymbol{\Sigma}_{n,0}                         & \boldsymbol{\Lambda}_{n,\boldsymbol{\theta}} \\[2ex]
    \boldsymbol{\Sigma}^{\T}_{n+1,n-1}                  & \boldsymbol{\Sigma}^{\T}_{n,n-1}                  & \boldsymbol{\Sigma}_{n-1,n-1}                       & \cdots & \boldsymbol{\Sigma}_{n-1,0}                       & \boldsymbol{\Lambda}_{n-1,\boldsymbol{\theta}} \\[2ex]
    \vdots                                            & \vdots                                            & \vdots                                              & \ddots & \vdots                                            & \vdots \\[2ex]
    \boldsymbol{\Sigma}^{\T}_{n,0}                    & \boldsymbol{\Sigma}^{\T}_{n,0}                    & \boldsymbol{\Sigma}^{\T}_{n-1,0}                    & \cdots & \boldsymbol{\Sigma}_{0,0}                         & \boldsymbol{\Lambda}_{0,\boldsymbol{\theta}} \\[2ex]
    \boldsymbol{\Lambda}^{\T}_{n,\boldsymbol{\theta}} & \boldsymbol{\Lambda}^{\T}_{n,\boldsymbol{\theta}} & \boldsymbol{\Lambda}^{\T}_{n-1,\boldsymbol{\theta}} & \cdots & \boldsymbol{\Lambda}^{\T}_{0,\boldsymbol{\theta}} & \boldsymbol{\Sigma}_{\boldsymbol{\theta},\boldsymbol{\theta}}
\end{bmatrix},
\end{equation}

where
\begin{align}
    \label{eqn_signp1np1}   \boldsymbol{\Sigma}_{n+1,n+1} & = \left(\left(\eye_{d} + \nabla_{\mathbf{x}} \mathbf{f}\Bigr\rvert_{n}  \Delta t \right) \boldsymbol{\Sigma}_{n,n} + \nabla^\T_{\boldsymbol{\theta}} \mathbf{f}\Bigr\rvert_{n} \boldsymbol{\Lambda}^{\T}_{\boldsymbol{n,\theta}} \, \Delta t \right)
\left(\eye_{d} + \nabla^{\T}_{\mathbf{x}} \mathbf{f}\Bigr\rvert_{n}  \Delta t \right) \\
\nonumber & \phantom{=} + \left(  \nabla_{\boldsymbol{\theta}} \mathbf{f}\Bigr\rvert_{n} \boldsymbol{\Sigma}_{\boldsymbol{\theta},\boldsymbol{\theta}} \, \Delta t
+ \left(\eye_{d} + \nabla^{\T}_{\mathbf{x}} \mathbf{f}\Bigr\rvert_{n}  \Delta t \right) \boldsymbol{\Lambda}_{n,\boldsymbol{\theta}} \right) \nabla^\T_{\boldsymbol{\theta}} \mathbf{f}\Bigr\rvert_{n} \, \Delta t,
\\[2ex]
\label{eqn_thetaterms} \boldsymbol{\Lambda}_{n+1,\boldsymbol{\theta}} & = \left(\eye_{d} + \nabla_{\mathbf{x}} \mathbf{f}\Bigr\rvert_{n}  \Delta t \right) \boldsymbol{\Lambda}_{n,\boldsymbol{\theta}}  + \nabla_{\boldsymbol{\theta}} \mathbf{f}\Bigr\rvert_{n} \boldsymbol{\Sigma}_{\boldsymbol{\theta},\boldsymbol{\theta}}  \, \Delta t.
\\[2ex]
\label{eqn_crossterms} \boldsymbol{\Sigma}_{n+1,\, j} & = \left(\eye_{d} + \nabla_{\mathbf{x}} \mathbf{f}\Bigr\rvert_{n} \Delta t \right)
\boldsymbol{\Sigma}_{n,\,j}  + \nabla_{\boldsymbol{\theta}} \mathbf{f}\Bigr\rvert_{n} \boldsymbol{\Lambda}^{\T}_{j,\boldsymbol{\theta}} \, \Delta t \qquad \text{for } 0\leq j \leq n,
\end{align}

If the above evolution of the covariance matrix can be related to the evolution of the sensitivity matrix, as presented in the section \ref{sec_dynamical_system_sensitivity} and equation \eqref{eqn_SSt_recur}, then the dependencies between the state vector and the parameters can be studied. This concept is developed in the next section.

\section{Relationship between sensitivity and forward propagation of uncertainty}
In this section the relationship between the evolution of the sensitivity matrix and the evolution of the covariance matrix of the joint distribution between all the state vectors until time $t_n$ and the parameters is developed.
Equation \eqref{eqn_signp1np1} can be expanded as follows
\begin{align}
    \label{eqn_signp1np1_simplified}   \boldsymbol{\Sigma}_{n+1,n+1} &=
    \phantom{+} \boldsymbol{\Sigma}_{n,n}
    &+& \, \nabla_{\mathbf{x}} \mathbf{f}\Bigr\rvert_{n} \boldsymbol{\Sigma}_{n,n} \, \Delta t
    &+& \,\nabla^\T_{\boldsymbol{\theta}} \mathbf{f}\Bigr\rvert_{n} \boldsymbol{\Lambda}^{\T}_{\boldsymbol{n,\theta}} \, \Delta t
    \nonumber \\ &\phantom{=}
    + \boldsymbol{\Sigma}_{n,n} \, \nabla^{\T}_{\mathbf{x}} \mathbf{f}\Bigr\rvert_{n}  \Delta t
    &+& \, \nabla_{\mathbf{x}} \mathbf{f}\Bigr\rvert_{n} \boldsymbol{\Sigma}_{n,n} \, \nabla^{\T}_{\mathbf{x}} \mathbf{f}\Bigr\rvert_{n} \,  \Delta t^2
    &+& \, \nabla^\T_{\boldsymbol{\theta}} \mathbf{f}\Bigr\rvert_{n} \boldsymbol{\Lambda}^{\T}_{\boldsymbol{n,\theta}} \, \nabla^{\T}_{\mathbf{x}} \mathbf{f}\Bigr\rvert_{n}  \, \Delta t^2
    \nonumber \\ &\phantom{=}
    + \nabla_{\boldsymbol{\theta}} \mathbf{f}\Bigr\rvert_{n} \boldsymbol{\Sigma}_{\boldsymbol{\theta},\boldsymbol{\theta}} \, \nabla^\T_{\boldsymbol{\theta}} \mathbf{f}\Bigr\rvert_{n} \, \Delta t^2
    &+& \,\, \boldsymbol{\Lambda}_{n,\boldsymbol{\theta}} \, \nabla^\T_{\boldsymbol{\theta}} \mathbf{f}\Bigr\rvert_{n} \, \Delta t
    &+& \, \nabla^\T_{\mathbf{x}} \mathbf{f}\Bigr\rvert_{n} \boldsymbol{\Lambda}_{n,\boldsymbol{\theta}} \, \nabla^\T_{\boldsymbol{\theta}} \mathbf{f}\Bigr\rvert_{n} \, \Delta t^2.
\end{align}
Assume the following
\begin{align}
    \label{eqn_ansatz_1}
    \boldsymbol{\Sigma}_{n,n} &= \mathbf{S}_{n} \mathbf{S}^{\T}_{n} \\
    \label{eqn_ansatz_2}
    \boldsymbol{\Lambda}_{n,\boldsymbol{\theta}} &= \mathbf{S}_{n}\\
    \label{eqn_ansatz_3}
    \boldsymbol{\Sigma}_{\boldsymbol{\theta},\boldsymbol{\theta}} &= \eye_{p}
\end{align}
Under the above assumptions, it can be deduced from equations \eqref{eqn_signp1np1_simplified} and \eqref{eqn_SSt_recur} that
\begin{equation}
    \label{eqn_signp1np1_substituted}
\boldsymbol{\Sigma}_{n+1,n+1} = \mathbf{S}_{n+1} \mathbf{S}^{\T}_{n+1}.
\end{equation}
Furthermore, equation \eqref{eqn_thetaterms} reads
\begin{align*}
\boldsymbol{\Lambda}_{n+1,\boldsymbol{\theta}} & = \left(\eye_{d} + \nabla_{\mathbf{x}} \mathbf{f}\Bigr\rvert_{n}  \Delta t \right) \mathbf{S}_n  + \nabla_{\boldsymbol{\theta}} \mathbf{f}\Bigr\rvert_{n} \, \Delta t,
\end{align*}
which, as evident from equation \eqref{eqn_linearised_sensitivity}, is the standard forward propagation of the sensitivity matrix. Hence
\begin{equation}
\label{eqn_thetaterms_substituted}     \boldsymbol{\Lambda}_{n+1,\boldsymbol{\theta}} = \mathbf{S}_{n+1}.
\end{equation}
Finally, the term $\boldsymbol{\Sigma}_{n+1,n}$ from equation \eqref{eqn_crossterms} can be written as
%
%
%
\begin{equation}
    \label{eqn_crossterms_substituted}
   \boldsymbol{\Sigma}_{n+1,\, n} =  \mathbf{S}_{n+1} \mathbf{S}^{\T}_{n}
\end{equation}

From equations \eqref{eqn_signp1np1_substituted}, \eqref{eqn_thetaterms_substituted}, and \eqref{eqn_crossterms_substituted}, it can be concluded that if the initial \emph{prior} uncertainty in $[\mathbf{x}^{\T}_{0}, \boldsymbol{\theta} ]^{\T}$ is assumed to be Gaussian with covariance
\begin{equation}
    \label{eqn_prior_cov}
    \cov \left(
    \begin{bmatrix}
        \mathbf{x}_0 \\[2ex]
        \boldsymbol{\theta}
    \end{bmatrix}
    \right) = \boldsymbol{\Sigma}_{0} =
    \begin{bmatrix}
        \mathbf{S}_{0} \mathbf{S}^{\T}_{0} & \mathbf{S}_0 \\[2ex]
        \mathbf{S}^{\T}_{0} & \eye_{p}
    \end{bmatrix},
\end{equation}
then the joint vector of $\boldsymbol{\theta}$, the parameters, and $[\mathbf{x}^{\T}_{n}, \mathbf{x}^{\T}_{n-1} , \ldots, \mathbf{x}^{\T}_{0}]^{\T}$, the state-vector corresponding to time instants $[t_0, t_1, \ldots, t_n]$, can be approximated, by considering only the first-order terms after linearisation, to be a Gaussian distribution with the following covariance
\begin{equation}
    \label{eqn_cov_joint_state_theta}
    \cov \left(
 \begin{bmatrix}
    \mathbf{x}_{n} \\[1ex]
    \mathbf{x}_{n-1} \\[1ex]
    \vdots \\[1ex]
    \mathbf{x}_{0} \\[1ex]
    \boldsymbol{\theta}
\end{bmatrix}
    \right) = \boldsymbol{\Sigma}_{n} =
\left[
    \begin{array}{ccccc|c}
         \mathbf{S}_{n} \mathbf{S}^{\T}_{n}   & \mathbf{S}_{n} \mathbf{S}^{\T}_{n-1}   & \mathbf{S}_{n} \mathbf{S}^{\T}_{n-2}   & \cdots & \mathbf{S}_{n} \mathbf{S}^{\T}_{0}   & \mathbf{S}_{n}    \\[2ex]
        \mathbf{S}_{n-1} \mathbf{S}^{\T}_{n} & \mathbf{S}_{n-1} \mathbf{S}^{\T}_{n-1} & \mathbf{S}_{n-1} \mathbf{S}^{\T}_{n-2} & \cdots & \mathbf{S}_{n-1} \mathbf{S}^{\T}_{0} & \mathbf{S}_{n-1}  \\[2ex]
        \mathbf{S}_{n-2} \mathbf{S}^{\T}_{n} & \mathbf{S}_{n-2} \mathbf{S}^{\T}_{n-1} & \mathbf{S}_{n-2} \mathbf{S}^{\T}_{n-2} & \cdots & \mathbf{S}_{n-2} \mathbf{S}^{\T}_{0} & \mathbf{S}_{n-2}  \\[2ex]
\vdots                               & \vdots                                 & \vdots                                 & \ddots & \vdots                        & \vdots  \\[2ex]
        \mathbf{S}_{0} \mathbf{S}^{\T}_{n} & \mathbf{S}_{0} \mathbf{S}^{\T}_{n-1} & \mathbf{S}_{0} \mathbf{S}^{\T}_{n-2} & \cdots & \mathbf{S}_{0} \mathbf{S}^{\T}_{0}         & \mathbf{S}_{0}    \\[2ex]
        \hline
\rule{0pt}{4ex}
\mathbf{S}^{\T}_n & \mathbf{S}^{\T}_{n-1} & \mathbf{S}^{\T}_{n-2} & \cdots & \mathbf{S}^{\T}_0 & \eye_p
\end{array}
\right].
\end{equation}

\begin{remark}
Note that a \emph{prior} mean for the vector $[\mathbf{x}^{\T}_{0}, \boldsymbol{\theta}^\T ]^{\T}$ is assumed to be
\begin{equation}
    \mean
    \left(
    \begin{bmatrix}
        \mathbf{x}_0 \\[2ex]
        \boldsymbol{\theta}
    \end{bmatrix}
    \right) = \boldsymbol{\mu}_{0} =
    \begin{bmatrix}
                \boldsymbol{\mu}_{\mathbf{x}_{0}} \\[1ex]
            \boldsymbol{\mu}_{\boldsymbol{\theta}}
    \end{bmatrix},
\end{equation}
based on which the mean vector of the state will propagate according to equation \eqref{eqn_discretised_forward_mean}, essentially according to the forward Euler method. While this propagated mean does not directly influence the \emph{posterior} uncertainty of the parameters, which depends only on the covariance matrix, it is important to note that the sensitivity terms in the covariance matrix of equation \eqref{eqn_cov_joint_state_theta} are evaluated at the propagated means. The propagated mean of the joint vector $[\mathbf{x}^{\T}_{n}, \mathbf{x}^{\T}_{n-1} , \ldots, \mathbf{x}^{\T}_{0}, \boldsymbol{\theta}^{\T} ]^{\T}$ is referred throughout this manuscript as $\boldsymbol{\mu}_n =[\boldsymbol{\mu}_{\mathbf{x}_n}^{\T}, \boldsymbol{\mu}_{\mathbf{x}_{n-1}}^{\T} , \ldots, \boldsymbol{\mu}_{\mathbf{x}_{0}}^{\T}, \boldsymbol{\mu}_{\boldsymbol{\theta}}^{\T} ]^{\T} $.
\end{remark}
\begin{remark}
    The required conditions presented in equations \eqref{eqn_ansatz_1}, \eqref{eqn_ansatz_2}, and \eqref{eqn_ansatz_3}, can also be derived without temporal discretisation of the sensitivity and linearised forward model. This is presented in Appendix \ref{app_differential_analysis}.
\end{remark}

\section{Measurements (observations)}
Having established how the covariance of the state and the parameters evolves in relation to the sensitivity matrix, the next task is to extend this framework to include the measurements. Eventually, one wants to obtain an expression for the joint distribution of the measuremennts and the parameters, so that conditioning this joint distribution on the measurements (implying that measurements are known) will yield information about how much can be learned about the parameters.

Consider a linear observation operator where $\mathbf{y}_n \in \mathbb{R}^m$ is measured at time $t_n$ according to
\begin{equation}
\label{eqn_observation_operator}
\mathbf{y}_n = \mathbf{H}_{n} \mathbf{x}_{n} + \boldsymbol{\epsilon}_{n},
\end{equation}
where $\mathbf{H}_n \in \mathbb{R}^{m\times d}$ is the observation operator at time $t_n$ and $\boldsymbol{\epsilon}_{n}$ is the measurement noise. Let $\boldsymbol{\epsilon}_{n}$ be independently (across all measurement times) distributed as
\begin{equation}
    \label{eqn_dist_epsilon}
\boldsymbol{\epsilon}_{n} \sim \mathcal{N}(\zero_{m}, \boldsymbol{\Upsilon}_{n}),
\end{equation}
where $\zero_{m}$ is a zero vector and $\boldsymbol{\Upsilon}_{n}$ is the covariance structure of the noise. From equations \eqref{eqn_cov_joint_state_theta} and \eqref{eqn_observation_operator}, it is easy to see that $[\mathbf{y}^{\T}_{n}, \mathbf{y}^{\T}_{n-1} , \ldots, \mathbf{y}^{\T}_{0},\boldsymbol{\theta}]^{\T}$ follows a Gaussian distribution with the following mean and covariance

\begin{equation}
    \label{eqn_cov_joint_obs_theta}
    \mean \left(
 \begin{bmatrix}
    \mathbf{y}_{n} \\[1ex]
    \mathbf{y}_{n-1} \\[1ex]
    \vdots \\[1ex]
    \mathbf{y}_{0} \\[1ex]
    \boldsymbol{\theta}
\end{bmatrix}
    \right)
    = \boldsymbol{\alpha}_{n}
    =
\begin{bmatrix}
   \mathbf{H}_{n} \,  \boldsymbol{\mu}_{\mathbf{x}_n} \\[1ex]
   \mathbf{H}_{n-1} \, \boldsymbol{\mu}_{\mathbf{x}_{n-1}} \\[1ex]
   \vdots \\[1ex]
 \mathbf{H}_{0} \boldsymbol{\mu}_{\mathbf{x}_{0}} \\[1ex]
 \boldsymbol{\mu}_{\boldsymbol{\theta}}
\end{bmatrix}
    \quad ; \quad
    \cov \left(
 \begin{bmatrix}
    \mathbf{y}_{n} \\[1ex]
    \mathbf{y}_{n-1} \\[1ex]
    \vdots \\[1ex]
    \mathbf{y}_{0} \\[1ex]
    \boldsymbol{\theta}
\end{bmatrix}
    \right)
    =
\begin{bmatrix}
    \mathbf{A}_n & \mathbf{B}_n \\[2ex]
    \mathbf{B}_n^{\T} & \eye_p
\end{bmatrix},
\end{equation}
where
\begin{equation}
    \setlength{\arraycolsep}{8pt}
    \label{eqn_def_An}
    \mathbf{A}_n =
    \begin{bmatrix}
\mathbf{H}_{n}  \,  \mathbf{S}_{n}  \,    \mathbf{S}^{\T}_{n} \, \mathbf{H}^{\T}_{n}   + \boldsymbol{\Upsilon}_n       & \mathbf{H}_{n}  \, \mathbf{S}_{n}  \, \mathbf{S}^{\T}_{n-1} \, \mathbf{H}^{\T}_{n-1}     & \cdots & \mathbf{H}_{n}  \, \mathbf{S}_{n}  \, \mathbf{S}^{\T}_{0} \, \mathbf{H}^{\T}_{0}      \\[2ex]
        \mathbf{H}_{n-1} \mathbf{S}_{n-1} \mathbf{S}^{\T}_{n}\, \mathbf{H}^{\T}_{n}         & \mathbf{H}_{n-1} \mathbf{S}_{n-1} \mathbf{S}^{\T}_{n-1} \mathbf{H}^{\T}_{n-1}  + \boldsymbol{\Upsilon}_{n-1} & \cdots & \mathbf{H}_{n-1} \mathbf{S}_{n-1} \mathbf{S}^{\T}_{0} \, \mathbf{H}^{\T}_{0}  \\[2ex]
        \vdots                                                  \,                             & \vdots                                                                           & \ddots & \vdots \\[2ex]
        \mathbf{H}_{0} \,    \mathbf{S}_{0} \,     \mathbf{S}^{\T}_{n}\, \mathbf{H}^{\T}_{n} & \mathbf{H}_{0} \, \mathbf{S}_{0} \, \mathbf{S}^{\T}_{n-1} \mathbf{H}^{\T}_{n-1}    & \cdots & \mathbf{H}_{0} \, \mathbf{S}_{0}  \, \mathbf{S}^{\T}_{0} \,  \mathbf{H}^{\T}_{0} + \boldsymbol{\Upsilon}_0      \\[2ex]
    \end{bmatrix}
\end{equation}
and
\begin{equation}
    \setlength{\arraycolsep}{22pt}
    \label{eqn_def_Bn}
    \mathbf{B}_n^{\T} =
    \begin{bmatrix}
\mathbf{S}^{\T}_{n} \, \mathbf{H}^{\T}_{n}                       & \mathbf{S}^{\T}_{n-1} \, \mathbf{H}^{\T}_{n-1}                         & \cdots & \mathbf{S}^{\T}_{0} \, \mathbf{H}^{\T}_{0}
    \end{bmatrix}.
\end{equation}

\begin{remark}
    A non-linear observation operator $\boldsymbol{\mathcal{H}}$ in equation \eqref{eqn_observation_operator}, as opposed to the linear operator $\mathbf{H}$, does not present any technical challenges to the formulation  as it can be linearised at the current mean values. Following this, in equations \eqref{eqn_def_An} and \eqref{eqn_def_Bn}, $\mathbf{H}$ would need to be replaced by the tangent operator $\nabla \boldsymbol{\mathcal{H}}\Bigr\rvert_{n}$.
\end{remark}

\section{Conditional distribution of the parameters}
The quantity of interest is the conditional distribution of parameters; \emph{i.e} how the beliefs about the parameters have changed from the \emph{prior} beliefs to the \emph{posterior} beliefs (the conditional distribution) by the measurements. More than the mean of the  conditional distribution, the covariance is of interest. This is due to two reasons: i) owing to the Gaussian approximations, the covariance entirely reflects the amount of uncertainty in the parameters; and ii) while the mean of the conditional distribution depends on the measurements, the  covariance does not. The latter is significant because \emph{a priori} the measurement values are not known. Consequently, the average (over all possible measurements) uncertainty in the parameters too is independent of the measurements, and hence can be studied in an \emph{a priori} manner.


From equation \eqref{eqn_cov_joint_obs_theta}, since the joint distribution of the parameter vector and the observables is Gaussian, the conditional distribution of the parameter vector given the measurements is also Gaussian and can be written as

\begin{equation}
    \label{eqn_cond_dist}
    p(\boldsymbol{\theta} \, | \, [\mathbf{y}^{\T}_{n}, \mathbf{y}^{\T}_{n-1} , \ldots, \mathbf{y}^{\T}_{0}]^{\T}) = \mathcal{N}(\boldsymbol{\beta}_n,\boldsymbol{\mathcal{C}}_n)
\end{equation}
with
\begin{equation}
    \label{eqn_cond_mean}
    \boldsymbol{\beta}_n = \boldsymbol{\mu}_{\boldsymbol{\theta}} + \mathbf{B}^{\T}_n \mathbf{A}^{-1}_n \left([\mathbf{y}^{o\T}_{n}, \mathbf{y}^{o\T}_{n-1} , \ldots, \mathbf{y}^{o\T}_{0}]^{\T} - [
            (\mathbf{H}_{n} \,  \boldsymbol{\mu}_{\mathbf{x}_n})^{\T},   (\mathbf{H}_{n-1} \, \boldsymbol{\mu}_{\mathbf{x}_{n-1}})^{\T},   \ldots,  (\mathbf{H}_{0} \boldsymbol{\mu}_{\mathbf{x}_{0}})^{\T}
    ]^{\T} \right)
\end{equation}
and
\begin{equation}
    \label{eqn_def_cond_cov}
    \boldsymbol{\mathcal{C}}_n = \eye_{p} -  \mathbf{B}^{\T}_n \mathbf{A}^{-1}_n \mathbf{B}_n,
\end{equation}
where $\mathbf{y}^o_i$ denotes the measurement value (the realisation of the random variable $\boldy_n$ observed) at $t_i$.
Note that the conditional covariance $\boldsymbol{\mathcal{C}}_n$ is independent of these measurement values $\mathbf{y}^o_i$. Furthermore, since the uncertainty in a Gaussian random variable, quantified by the differential entropy, depends only on the covariance matrix, the posterior distribution uncertainty does not depend on the measurements.

\section{Asymptotic analysis of the conditional covariance}
\label{sec_asympotic_analysis}
In this section, the behaviour of the conditional covariance matrix $\boldsymbol{\mathcal{C}}_n$ as $n\to \infty$ is considered. From equation \eqref{eqn_def_An} $\mathbf{A}_n$ can be written as
\begin{equation}
    \mathbf{A}_n = \mathbf{B}_n \, \mathbf{B}^{\T}_n + \boldsymbol{\Gamma}_n \;,
\end{equation}
where $\boldsymbol{\Gamma}_n$ is a diagonal matrix with elements as follows
\begin{align}
    \label{eqn_def_Gamma}
    \Gamma_{i,i} &= \boldsymbol{\Upsilon}_i  \qquad ; \quad 0 \leq i \leq n.
\end{align}
By applying the Kailath variant of the Sherman-Morrison-Woodbury identity the inverse of $\mathbf{A}_n$ can be expanded as
\begin{equation}
    \label{eqn_inv_A}
    \mathbf{A}^{-1}_n = \boldsymbol{\Gamma}^{-1}_n - \boldsymbol{\Gamma}^{-1}_n \, \mathbf{B}_n \left( \eye_p + \, \mathbf{B}^{\T}_n \, \boldsymbol{\Gamma}^{-1}_n \,  \mathbf{B}_n \right)^{-1}
    \mathbf{B}^{\T}_n \, \boldsymbol{\Gamma}^{-1}_n.
\end{equation}
Plugging this in equation \eqref{eqn_def_cond_cov} yields

\begin{align}
    \boldsymbol{\mathcal{C}}_n &= \eye_{p} -
    \mathbf{B}^{\T}_n \boldsymbol{\Gamma}^{-1}_n \mathbf{B}_n
    + \mathbf{B}^{\T}_n \,\boldsymbol{\Gamma}^{-1}_n \,
    \mathbf{B}_n \left( \eye_p + \mathbf{B}^{\T}_n  \, \boldsymbol{\Gamma}^{-1}_n \, \mathbf{B}_n \right)^{-1}
\mathbf{B}^{\T}_n \, \boldsymbol{\Gamma}^{-1}_n \mathbf{B}_n \\
&= \eye_{p} -
\mathbf{D}_n
    + \mathbf{D}_n \left( \eye_p + \mathbf{D}_n \right)^{-1}
    \mathbf{D}_n, \label{eqn_Cn_preSVD}
\end{align}
where
\begin{equation}
    \label{eqn_def_D}
    \mathbf{D}_n = \mathbf{B}^{\T}_n  \, \boldsymbol{\Gamma}^{-1}_n \, \mathbf{B}_n .
\end{equation}

The matrix $\mathbf{D}$ is symmetric, and can be factorised by singular value decomposition (SVD) as follows 
\begin{equation}
    \label{eqn_svd}
    \mathbf{D}_n = \mathbf{U}_n \boldsymbol{\Phi}_n \mathbf{U}_n^{\T} ,
\end{equation}
with
\begin{equation}
    \mathbf{U}_n  \mathbf{U}_n^{\T} = \eye_p ,
\end{equation}
and $\boldsymbol{\Phi}_n$ is a diagonal matrix with diagonal entries equal to the eigenvalues, $\lambda_i$, of $\mathbf{D}_n$.
\begin{equation}
    \Phi_{n_{i,i}} = \lambda_i.
\end{equation}
Due to the symmetric nature of $\mathbf{D}_n$, all the eigenvalues are real. Furthermore, if $\mathbf{D}_n$ is positive-definite then all eigenvalues are positive. Substituting $\mathbf{D}_n$ from equation \eqref{eqn_svd} in equation \eqref{eqn_Cn_preSVD} yields
\begin{align}
    \label{eqn_Cn_asymptotic}
    \boldsymbol{\mathcal{C}}_n
    &=
    \eye_{p} -
\mathbf{U}_n \boldsymbol{\Phi}_n \mathbf{U}_n^{\T}
    + \mathbf{U}_n \boldsymbol{\Phi}_n \mathbf{U}_n^{\T} \left( \eye_p + \mathbf{U}_n \boldsymbol{\Phi}_n \mathbf{U}_n^{\T} \right)^{-1}
    \mathbf{U}_n \boldsymbol{\Phi}_n \mathbf{U}_n^{\T} \\
    &=
    \eye_{p} -
\mathbf{U}_n \boldsymbol{\Phi}_n \mathbf{U}_n^{\T}
    + \mathbf{U}_n \boldsymbol{\Phi}_n \mathbf{U}_n^{\T} \left( \mathbf{U}_n \mathbf{U}_n^{\T} + \mathbf{U}_n \boldsymbol{\Phi}_n \mathbf{U}_n^{\T} \right)^{-1}
    \mathbf{U}_n \boldsymbol{\Phi}_n \mathbf{U}_n^{\T} \\
    &=
    \eye_{p} -
\mathbf{U}_n \boldsymbol{\Phi}_n \mathbf{U}_n^{\T}
    + \mathbf{U}_n \boldsymbol{\Phi}_n \mathbf{U}_n^{\T}
    \left( \mathbf{U}_n (\eye_p + \boldsymbol{\Phi}_n) \mathbf{U}_n^{\T} \right)^{-1}
    \mathbf{U}_n \boldsymbol{\Phi}_n \mathbf{U}_n^{\T} \\
    &=
    \mathbf{U}_n \mathbf{U}_n^{\T} -
\mathbf{U}_n \boldsymbol{\Phi}_n \mathbf{U}_n^{\T}
    + \mathbf{U}_n \boldsymbol{\Phi}_n
    \left( \eye_p + \boldsymbol{\Phi}_n \right)^{-1}
    \boldsymbol{\Phi}_n \mathbf{U}_n^{\T} \\[1ex]
    &=
    \mathbf{U}_n
    \underbrace{\left[ \eye_{p} - \boldsymbol{\Phi}_n + \boldsymbol{\Phi}_n (\eye_p + \boldsymbol{\Phi}_n)^{-1} \boldsymbol{\Phi}_n \right]}_{\textstyle\mathbf{P}_n}
    \mathbf{U}_n^{\T}. \label{eqn_Cn_asymptotic_last}
\end{align}

In the above $\mathbf{P}_n$ is a diagonal matrix with the entries
\begin{equation}
    \label{eqn_Pn_entries}
    P_{n_{i,i}} = 1-\lambda_i + \frac{\lambda_i^2}{1+\lambda_i}
    = \frac{1}{1+\lambda_i}.
\end{equation}

If the minimum eigenvalue of $\mathbf{D}_n$ is much larger than 1, \emph{i.e.}
\begin{equation}
\min_i \lambda_i \gg 1,
\end{equation}
then
\begin{equation}
    P_{n_{i,i}} \approx \frac{1}{\lambda_i},
\end{equation}
and
\begin{equation}
    \mathbf{P}_n \approx \boldsymbol{\Phi}^{-1}.
\end{equation}
Consequently, equation \eqref{eqn_Cn_asymptotic_last} yields
\begin{equation}
    \boldsymbol{\mathcal{C}}_n \approx \mathbf{U}_n
    \boldsymbol{\Phi}^{-1}
    \mathbf{U}_n^{\T}
    = \mathbf{D}^{-1}_n.
\end{equation}
Finally, from the above and equations \eqref{eqn_def_D}, \eqref{eqn_def_Gamma}, and \eqref{eqn_def_Bn}, the conditional covariance matrix can be written as
\begin{equation}
    \label{eqn_Cn_final}
    \boldsymbol{\mathcal{C}}_n \approx
    \left(
    \sum^{n}_{i=0} \left( \mathbf{S}^{\T}_{i} \, \mathbf{H}^{\T}_{i} \,
        \boldsymbol{\Upsilon}^{-1}_i \,
 \mathbf{H}_{i} \, \mathbf{S}_{i} \right)
\right)^{-1}.
\end{equation}

It can hence be concluded that if the minimum eigenvalue of $\mathbf{D}_n$ monotonically increases as $n$ increases then
\begin{equation}
    \label{eqn_Cn_limit}
    \lim_{n\to \infty} \boldsymbol{\mathcal{C}}_n =
    \left(
    \sum^{n}_{i=0} \left( \mathbf{S}^{\T}_{i} \, \mathbf{H}^{\T}_{i} \,
        \boldsymbol{\Upsilon}^{-1}_i \,
 \mathbf{H}_{i} \, \mathbf{S}_{i} \right)
\right)^{-1}.
\end{equation}
Let the eigenvalues of $\mathbf{D}_n$ be denoted in decreasing order as $\lambda_1(\mathbf{D}_n) \geq \lambda_2(\mathbf{D}_n) \geq \ldots \geq \lambda_p(\mathbf{D}_n)$. The behaviour of the minimum eigenvalue of $\lambda_p(\mathbf{D}_n)$ is of concern. Note that $\mathbf{D}_n$ can be written as
\begin{equation}
    \label{eqn_Dn_expanded}
    \mathbf{D}_n = \sum^{n}_{i=0} \left( \mathbf{S}^{\T}_{i} \, \mathbf{H}^{\T}_{i} \,
        \boldsymbol{\Upsilon}^{-1}_i \,
\mathbf{H}_{i} \, \mathbf{S}_{i} \right)
\end{equation}
Consequently,
\begin{equation}
    \label{eqn_Dsum}
    \mathbf{D}_{n+1} = \mathbf{D}_{n} +
    \underbrace{
        \mathbf{S}^{\T}_{n+1} \, \mathbf{H}^{\T}_{n+1} \,
    \boldsymbol{\Upsilon}^{-1}_{n+1}\, \mathbf{H}_{n+1} \, \mathbf{S}_{n+1}
}_{\textstyle \mathbf{Q}_{n+1}}.
\end{equation}
$\mathbf{D}_n$ and $\mathbf{Q}_n$ are both symmetric matrices. Let the eigenvalues of $\mathbf{Q}_{n+1}$ be denoted in decreasing order as $\lambda_1(\mathbf{Q}_{n+1}) \geq \lambda_2(\mathbf{Q}_{n+1}) \geq \ldots \geq \lambda_p(\mathbf{Q}_{n+1})$. From equation \eqref{eqn_Dsum} one has
\begin{equation}
    \label{eqn_trace}
    \mathrm{Tr}\,(\mathbf{D}_{n+1}) = \mathrm{Tr}\,(\mathbf{D}_{n}) + \mathrm{Tr}\,(\mathbf{Q}_{n+1}),
\end{equation}
where $\mathrm{Tr}$ denotes the trace. Expressed in terms of the eigenvalues of the respective matrices, the above reads
\begin{equation}
    \label{eqn_trace_eigen}
    \sum_{i=1}^{p} \lambda_i(\mathbf{D}_{n+1}) =
    \sum_{i=1}^{p} \lambda_i(\mathbf{D}_{n}) +
    \sum_{i=1}^{p} \lambda_i(\mathbf{Q}_{n+1}).
\end{equation}
From several inequalities on the sums of eigenvalues of Hermitian matrices, specifically the \emph{Ky Fan inequality} \cite{fulton2000eigenvalues,fan1949theorem}, one has
\begin{equation}
    \label{eqn_fan}
    \sum_{i=1}^{r} \lambda_i(\mathbf{D}_{n+1}) \leq
    \sum_{i=1}^{r} \lambda_i(\mathbf{D}_{n}) +
    \sum_{i=1}^{r} \lambda_i(\mathbf{Q}_{n+1}) \qquad ; \quad r \leq p.
\end{equation}
Substituting $r={p-1}$ in equation \eqref{eqn_fan} and subtracting it from equation \eqref{eqn_trace_eigen} results in
\begin{equation}
    \label{eqn_ineq}
    \lambda_p(\mathbf{D}_{n+1}) \geq
    \lambda_p(\mathbf{D}_{n}) +
    \lambda_p(\mathbf{Q}_{n+1}).
\end{equation}
Consequently, if $\mathbf{Q}_{n+1}$ is full rank then $\lambda_p(\mathbf{Q}_{n+1})>0$ and
\begin{equation}
    \label{eqn_ineq_final}
    \lambda_p(\mathbf{D}_{n+1}) >
    \lambda_p(\mathbf{D}_{n}),
\end{equation}
which implies that the minimum eigenvalue of $\mathbf{D}_n$ is monotonically increasing.

The above results are put in the perspective of classical non-linear regression analysis \cite{seber1989nonlinear,banks2009inverse} by assuming that the observation operator $\mathbf{H}_i$ is equal to identity for all $i$. Then, under a further assumption that $\mathbf{Q}_i =  \left(\nabla^\T_{\boldsymbol{\theta}} \mathbf{x}\Bigr\rvert_{i}\right) \boldsymbol{\Upsilon}_i^{-1} \left(\nabla_{\boldsymbol{\theta}} \mathbf{x}\Bigr\rvert_{i} \right)$ is full-rank for all $i$, the conditional covariance matrix of the parameter is
\begin{equation}
    \label{eqn_Cn_simplified}
        \lim_{n\to \infty} \boldsymbol{\mathcal{C}}_n =
        \boldsymbol{\mathcal{M}}^{-1},
\end{equation}
where $\boldsymbol{\mathcal{M}}$ is the Fisher information matrix defined as
\begin{equation}
    \label{eqn_fim}
    \boldsymbol{\mathcal{M}} =
    \sum^{n}_{i=0}  \left[ \left(\nabla^\T_{\boldsymbol{\theta}} \mathbf{x}\Bigr\rvert_{i}\right) \boldsymbol{\Upsilon}_i^{-1} \left(\nabla_{\boldsymbol{\theta}} \mathbf{x}\Bigr\rvert_{i} \right) \right].
\end{equation}

\section{Conditional covariance for finite $n$}
\label{sec:conditional_covariance_for_finite_n}
From equation \eqref{eqn_Cn_asymptotic_last} we have
\begin{equation}
    \label{eqn_Cn_eigForm}
    \boldsymbol{\mathcal{C}}_n = \mathbf{U}_n \, \mathbf{P}_n \, \mathbf{U}_n^\T,
\end{equation}

where $\mathbf{P}_n$ is given by equation \eqref{eqn_Pn_entries}. Consider the matrix $(\mathbf{D}_n + \eye_p)^{-1}$, it can be expanded as
\begin{align}
    (\mathbf{D}_n + \eye_p)^{-1}    & = \left(\mathbf{U}_n \boldsymbol{\Phi}_n \mathbf{U}_n^\T + \eye_p \right)^{-1}\\
                                    & = \left(\mathbf{U}_n \boldsymbol{\Phi}_n \mathbf{U}_n^\T + \mathbf{U}_n \mathbf{U}_n^\T \right)^{-1}\\
                                    & = \left(\mathbf{U}_n (\boldsymbol{\Phi}_n + \eye_p) \mathbf{U}_n^\T \right)^{-1}\\
                                    & = \;\mathbf{U}_n (\boldsymbol{\Phi}_n + \eye_p)^{-1} \mathbf{U}_n^\T \\
                                    & = \; \mathbf{U}_n \, \mathbf{P}_n \, \mathbf{U}_n^\T \;=\; \boldsymbol{\mathcal{C}}_n.
\end{align}

Following the above and equation \eqref{eqn_Dn_expanded}, the conditional covariance matrix can be written as
\begin{equation}
\label{eqn_Cn_final_computable}
\boldsymbol{\mathcal{C}}_n = \left(\eye_p + \sum^{n}_{i=0} \left( \mathbf{S}^{\T}_{i} \, \mathbf{H}^{\T}_{i} \,
        \boldsymbol{\Upsilon}^{-1}_i \,
 \mathbf{H}_{i} \, \mathbf{S}_{i} \right) \right)^{-1}.
\end{equation}

\section{Information gain}
\label{sec_inforamtion_gain}
In this section the gain in information about the parameters by the measurements is considered. For details of such an information-theoretic approach the reader is referred to \cite{pant2015Information}. The gain in information about the parameter vector $\boldsymbol{\theta}$ by the measurements of $\mathbf{z}_n = [\mathbf{y}^{\T}_{n}, \mathbf{y}^{\T}_{n-1} , \ldots, \mathbf{y}^{\T}_{0}]^{\T}$ is given by the mutual information between $\mathbf{z}_n$ and $\boldsymbol{\theta}$, which is equal to the difference between the differential entropies of the prior distribution $p(\boldsymbol{\theta})$ and the conditional distribution $p(\boldsymbol{\theta}|\mathbf{z}_n)$. From equations \eqref{eqn_prior_cov}, \eqref{eqn_cond_dist}, and \eqref{eqn_def_cond_cov}, this gain in information can be written as 

\begin{equation}
\label{eqn_gain_joint}
\mathcal{I}_n = \frac12 \ln \left[ \det(\eye_p)\right]  - \frac12 \ln \left[ \det (\boldsymbol{\mathcal{C}}_n) \right] = -\frac12 \ln \left[ \det (\boldsymbol{\mathcal{C}}_n) \right],
\end{equation}
where $\det(\cdot)$ denotes the determinant. The above can be expanded through equation \eqref{eqn_Cn_final_computable} as
\begin{equation}
\label{eqn_gain_joint_computable}
\mathcal{I}_n =  \frac12 \ln \left[ \det \left(\eye_p + \sum^{n}_{i=0} \left( \mathbf{S}^{\T}_{i} \, \mathbf{H}^{\T}_{i} \,
        \boldsymbol{\Upsilon}^{-1}_i \,
\mathbf{H}_{i} \, \mathbf{S}_{i} \right) \right) \right].
\end{equation}


Note that the above represents the information gain for the joint vector of all the parameters. Commonly, one is interested in individual parameters, for which the information gain is now presented.
Let $\boldsymbol{\theta}^ {\left\{ \mathcal{S} \right\}}$ denote the vector of a subset of parameters indexed by the elements of set $\mathcal{S}$ and $\boldsymbol{\theta}^ {\left\{ \mathcal{S}^\complement \right\}}$ denote the vector of the remaining parameters, the complement of set $\mathcal{S}$. Hence, $\boldsymbol{\theta}^{\left\{ i \right\}}$ denotes the $i^{th}$ parameter, $\boldsymbol{\theta}^{\left\{ i,j \right\}}$ denotes the vector formed by taking the $i^{th}$ and $j^{th}$ parameters, and so on. The conditional covariance matrix $\ccn$ can be decomposed into the components of $\thetass$ and $\thetassc$ as 

\begin{equation}
    \ccn = 
        \left(
    \begin{bmatrix}
        \eye_\cards & \zero \\[2ex]
        \zero & \eye_{p-\cards}
    \end{bmatrix}
        +
        \sum^{n}_{i=0}
    \begin{bmatrix}
         \left( \ssst_{i} \, \mathbf{H}^{\T}_{i} \,
        \boldsymbol{\Upsilon}^{-1}_i \,
        \mathbf{H}_{i} \, \sss_{i} \right)
        &
         \left( \ssst_{i} \, \mathbf{H}^{\T}_{i} \,
        \boldsymbol{\Upsilon}^{-1}_i \,
        \mathbf{H}_{i} \, \sssc_{i} \right)
        \\[2ex]
         \left( \sssct_{i} \, \mathbf{H}^{\T}_{i} \,
        \boldsymbol{\Upsilon}^{-1}_i \,
        \mathbf{H}_{i} \, \sss_{i} \right)
         &
 \left( \sssct_{i} \, \mathbf{H}^{\T}_{i} \,
        \boldsymbol{\Upsilon}^{-1}_i \,
 \mathbf{H}_{i} \, \sssc_{i} \right)
    \end{bmatrix}
        \right)^{-1},
\end{equation}

\begin{equation}
    \label{eqn_split_2}
\begin{split}
    \ccn & =  
        \left(
    \begin{bmatrix}
        \eye_\cards & \zero \\[2ex]
        \zero & \eye_{p-\cards}
    \end{bmatrix}
        +
    \begin{bmatrix}
        \overbrace{
        \sum^{n}_{i=0}
         \left( \ssst_{i} \, \mathbf{H}^{\T}_{i} \,
        \boldsymbol{\Upsilon}^{-1}_i \,
        \mathbf{H}_{i} \, \sss_{i} \right)
    }^{\mathbf{D}_n^\mathcal{\{S\}}}
        &
        \overbrace{
        \sum^{n}_{i=0}
         \left( \ssst_{i} \, \mathbf{H}^{\T}_{i} \,
        \boldsymbol{\Upsilon}^{-1}_i \,
        \mathbf{H}_{i} \, \sssc_{i} \right)
    }^{\mathbf{D}_n^{\{\mathcal{S,S}^\complement\}}}
        \\[2ex]
        \underbrace{
        \sum^{n}_{i=0}
         \left( \sssct_{i} \, \mathbf{H}^{\T}_{i} \,
        \boldsymbol{\Upsilon}^{-1}_i \,
        \mathbf{H}_{i} \, \sss_{i} \right)
    }_{\mathbf{D}_n^{\{\mathcal{S}^\complement\!,\mathcal{S}\}}}
         &
        \underbrace{
        \sum^{n}_{i=0}
 \left( \sssct_{i} \, \mathbf{H}^{\T}_{i} \,
        \boldsymbol{\Upsilon}^{-1}_i \,
 \mathbf{H}_{i} \, \sssc_{i} \right)
    }_{\mathbf{D}_n^{\{\mathcal{S}^\complement\}}}
    \end{bmatrix}
        \right)^{-1}\\
        & = 
    \begin{bmatrix}
        \eye_\cards +  {\mathbf{D}_n^\mathcal{\{S\}}} & 
        {\mathbf{D}_n^{\{\mathcal{S,S}^\complement\}}} \\[2ex]
        {\mathbf{D}_n^{\{\mathcal{S}^\complement\!,\mathcal{S}\}}}&
        \eye_{p-\cards} + {\mathbf{D}_n^{\{\mathcal{S}^\complement\}}}
    \end{bmatrix}^{-1},\\
\end{split}
\end{equation}

where $\mathsfit{s}$  is the cardinality of $\mathcal{S}$, and $\sss \in \mathbf{R}^{d\times \cards}$ and $\sssc \in \mathbf{R}^{d\times (p-\cards)}$ are the sensitivity matrices for $\thetass$ and $\thetassc$, respectively, i.e. $\sss = \delthetass$ and $\sssc = \delthetassc$. Given the above decomposition, the marginal covariance of $\thetass$ given the measurements can be written as the Schur complement of the matrix $ \left[\eye_{p-\cards} + \sum^{n}_{i=0} \left( \sssct_{i} \, \mathbf{H}^{\T}_{i} \,
        \boldsymbol{\Upsilon}^{-1}_i \,
 \mathbf{H}_{i} \, \sssc_{i} \right) \right]$
 in $\ccn$
 as follows
 \begin{equation}
     \label{eqn_final_marginal_variance}
     \ccns = 
     \left[
     \left( \eye_\cards +  {\mathbf{D}_n^\mathcal{\{S\}}} \right)
     - \left({\mathbf{D}_n^{\{\mathcal{S,S}^\complement\}}} \right)
    \left( \eye_{p-\cards} + {\mathbf{D}_n^{\{\mathcal{S}^\complement\}}} \right)^{-1}
\left({\mathbf{D}_n^{\{\mathcal{S}^\complement,\mathcal{S}\}}} \right)
     \right]^{-1},
 \end{equation}

and the information gain $\inns$ as

\begin{equation}
\label{eqn_gain_marginal}
    \inns =  -\frac12 \ln \left[ \det (\ccns) \right] = \frac12 \ln \left[
     \left( \eye_\cards +  {\mathbf{D}_n^\mathcal{\{S\}}} \right)
     - \left({\mathbf{D}_n^{\{\mathcal{S,S}^\complement\}}} \right)
    \left( \eye_{p-\cards} + {\mathbf{D}_n^{\{\mathcal{S}^\complement\}}} \right)^{-1}
\left({\mathbf{D}_n^{\{\mathcal{S}^\complement\!,\mathcal{S}\}}} \right)
     \right].
\end{equation}

Another quantity of interest is the correlation between two subsets of parameters $\bthetas$ and $\bthetaw$. In an information-theoretic context this can be assessed by how much more information is gained about the parameters $\bthetas$ in addition to $\inns$ if $\bthetaw$ was also known, i.e. the mutual information between $\bthetas$ and $\bthetaw$ given the measurements. Similar to the procedure employed in equation \eqref{eqn_split_2}, by splitting $\ccn$ into three components for $\bthetas$, $\bthetaw$, and $\bthetaswc$, one can write the conditional covariance $\ccnsgw$ of the parameters $\bthetas$ given the measurements and, additionally, the parameters $\bthetaw$ as follows

\begin{equation}
    \label{eqn_cond_sgp}
\ccnsgw = 
     \left[
     \left( \eye_\cards +  {\mathbf{D}_n^\mathcal{\{S\}}} \right)
     - \left({\mathbf{D}_n^{\{\mathcal{S},\suwc\}}} \right)
    \left( \eye_{p-\cards-\cardw} + {\mathbf{D}_n^{\{\suwc\}}} \right)^{-1}
\left({\mathbf{D}_n^{\{\suwc,\mathcal{S}\}}} \right)
     \right]^{-1},
\end{equation}

\noindent
where $\cardw$ is the cardinality of $\mathcal{W}$,
\begin{equation}
    {\mathbf{D}_n^{\{\suwc\}}} = 
    \sum^{n}_{i=0}
 \left( \sssuwct_{i} \, \mathbf{H}^{\T}_{i} \,
        \boldsymbol{\Upsilon}^{-1}_i \,
    \mathbf{H}_{i} \, \sssuwc_{i} \right) \quad \text{with } \sssuwc = \delthetassuwc,
\end{equation}
\noindent
and
\begin{equation}
    {\mathbf{D}_n^{\{\mathcal{S},\suwc\}}} = 
    \sum^{n}_{i=0}
 \left( \ssst_{i} \, \mathbf{H}^{\T}_{i} \,
        \boldsymbol{\Upsilon}^{-1}_i \,
    \mathbf{H}_{i} \, \sssuwc_{i} \right)
\end{equation}

The information gain $\innsgw$ about the parameters $\bthetas$ given both the measurements and the parameters $\bthetaw$ is

\begin{equation}
\label{eqn_gain_conditional}
    \begin{split}
        \innsgw & =  -\frac12 \ln \left[ \det (\ccnsgw) \right] \\
        & = \frac12 \ln \left[
     \left( \eye_\cards +  {\mathbf{D}_n^\mathcal{\{S\}}} \right)
     - \left({\mathbf{D}_n^{\{\mathcal{S},\suwc\}}} \right)
    \left( \eye_{p-\cards-\cardw} + {\mathbf{D}_n^{\{\suwc\}}} \right)^{-1}
\left({\mathbf{D}_n^{\{\suwc,\mathcal{S}\}}} \right)
     \right].
    \end{split}
\end{equation}

Lastly, the conditional mutual information, i.e. the additional (after the measurements are known) information gained about the parameters $\bthetas$ due to the knowledge of $\bthetaw$ is
\begin{equation}
    \label{eqn_mi}
    \mathcal{I}_n^{\{\mathcal{S};\mathcal{W}\}} = \innsgw - \inns
\end{equation}

\begin{remark}
    $\inns$ is the gain in information about the parameters $\bthetas$ given the measurements and when nothing is known about the parameters $\bthetasc$.
\end{remark}

\begin{remark}
    $\innsgw$ is the gain in information about the parameters $\bthetas$ given the measurements and the parameters $\bthetaw$, when nothing is known about the parameters $\bthetaswc$.
\end{remark}


\begin{remark}
In \cite{pant2015Information}, the authors suggested a method to interpret the information gains $\inns$ and $\innsgw$ when the set $\mathcal{S}$ contained a single parameter by proposing a hypothetical measurement device. This is not necessary in the current formulation as all the distributions are approximated to be Gaussian. Therefore, when $\mathcal{S}$ contains only a single parameter, the conditional covariances $\ccns$ and $\ccnsgw$ are scalar quantities representing the posterior variances of the parameter $\bthetas$. When $\mathcal{S}$ is contains more than one parameter, the quantities $\inns$ and $\innsgw$ are scalars that quantify the gains in information.
\end{remark}

The above developed functions for information gains (and associated variances) are collectively referred as `Information Sensitivity Functions'. From this point onwards, the terms \emph{marginal posterior variance} or just \emph{marginal variance} for a parameter subset $\bthetas$ refers to the the variance conditioned on only the measurements, equation \eqref{eqn_final_marginal_variance}, and the corresponding information gain, equation \eqref{eqn_gain_marginal}, is referred as the \emph{marginal information gain}. Similarly the term \emph{conditional variance} is used to refer to the variance when the measurements and additionally a parameter subset $\bthetaw$ is given, equation \eqref{eqn_cond_sgp}, and the corresponding information gain is referred as the \emph{conditional information gain}, equation \eqref{eqn_gain_conditional}. Lastly the information shared between two subsets of parameters given the measurements, equation \eqref{eqn_mi}, is referred as the \emph{conditional mutual information} or just the \emph{mutual information}. Finally the vector $\mathbf{z}_n = [\mathbf{y}^{\T}_{n}, \mathbf{y}^{\T}_{n-1} , \ldots, \mathbf{y}^{\T}_{0}]^{\T}$ is used to denote a collection of all measurement vectors up to time $t_n$.


\section{Results and discussion}
In this section, the above developed theory is applied to study three dynamical systems.

\label{sec_results_and_discussion}

\subsection{Three-element Windkessel model}
\begin{figure}[httb]
\centering
    \subfloat[Schematic of a three-element Windkessel model.\label{fig_RCR_schematic}]{%
    \raisebox{0.8cm}{
 \begin{circuitikz}[scale=0.74, transform shape] \draw
(0,0) node[anchor=east]{$P^i$}
to[R, l=$R_p$, i>_=$q^i$,*-*] (3.0,0)
to[R, l=$R_d$, *-*, i=$q^o$] (6.0,0.0) 
(3.0,0) to [C, l=$C$, i=$q^i-q^o$, *-*] (3,-2.0)
(6.0,0.0)  node[anchor=west]{$P_{\mathrm{ven}}=0$}
(3,-2.0) node[anchor=north]{$P_{\mathrm{ext}}=0$}
(3,0) node[anchor=south]{$P^c$}
;\end{circuitikz}
}
    }
    \quad
    \subfloat[Inlet flow-rate curve and Windkessel pressure solution, see equation \eqref{eqn_RCR}, with nominal parameter values. \label{fig_RCR_pq}]{%
      \includegraphics[width=0.55\textwidth]{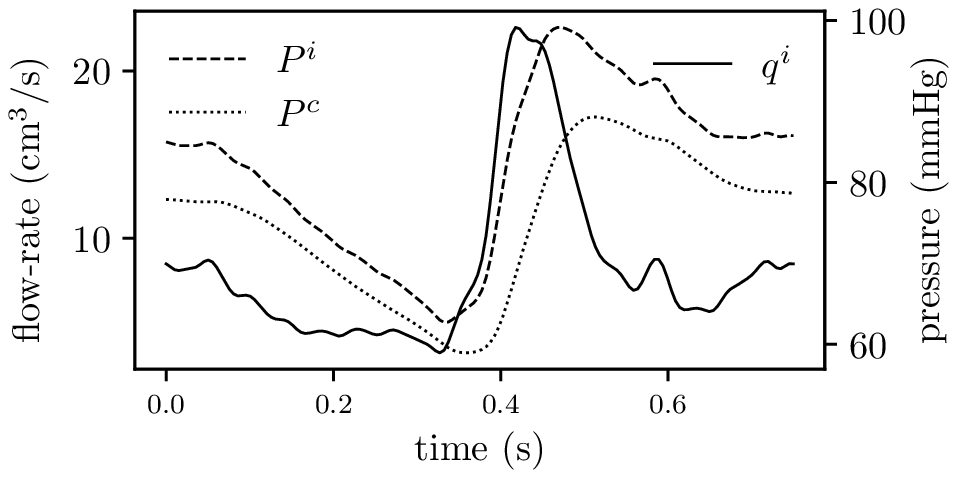}
    }
    \caption{The three-element Windkessel model, flow-rate curve utilised, and pressure solutions.}
\end{figure}

\begin{figure}[htpb]
    \centering
    \includegraphics[width=1.0\linewidth]{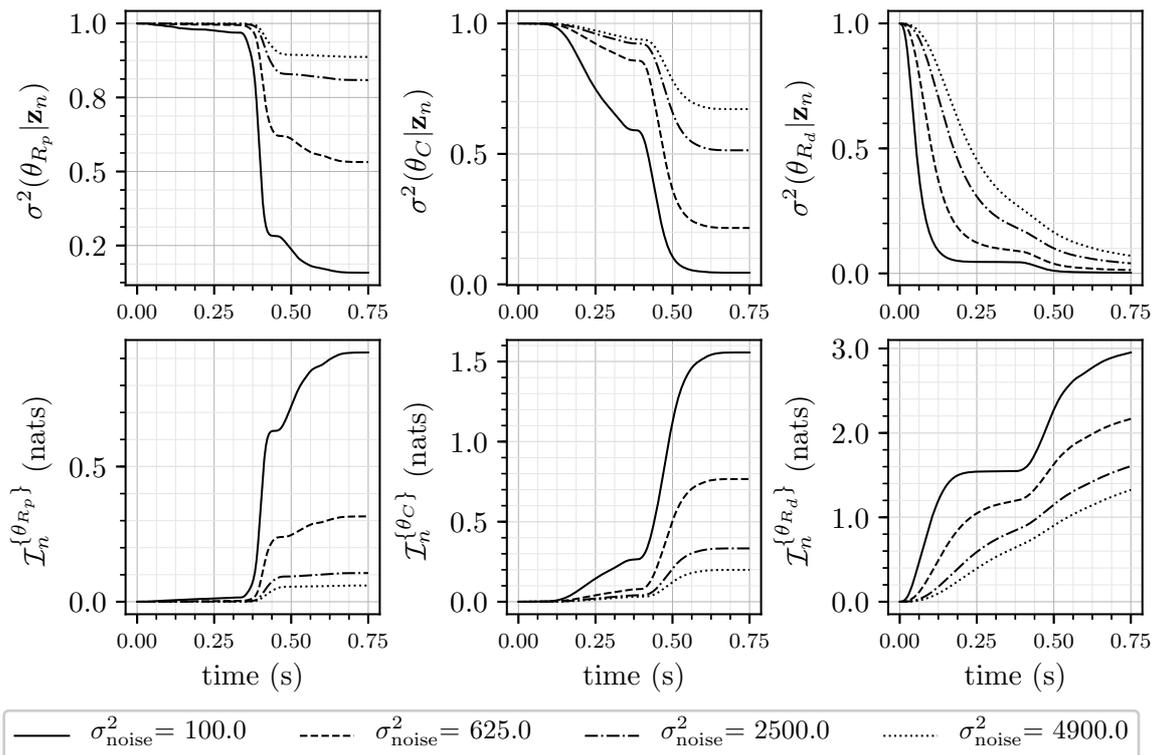}
    \caption{Marginal posterior variances (top row) and marginal information gains (bottom row) for the three Windkessel model parameters at four different levels of measurement noise.}
    \label{fig_RCR_marginal_var_isf}
\end{figure}

Windkessel models are widely used to describe arterial haemodynamics \cite{westerhof2009arterial}. Increasingly, they are also being used as boundary conditions in 3D computational fluid dynamics simulations to assess patient-specific behaviour \cite{vignon2006outflow,pant2014methodological}. In order to perform patient-specific analysis, it is imperative that the parameters of the Windkessel model are estimated from measurements taken in each patient individually. A three-element Windkessel models is shown in Figure \ref{fig_RCR_schematic} and consists of three parameters: $R_p$ (proximal resistance) which represents the hydraulic resistance of large vessels; $C$ (capacitance) which represents the compliance of large vessels; and $R_d$ which represents the resistance of small vessels in the microcirculation. Note that these models utilise the electric analogy to fluid flow where pressure $P$ is seen as voltage and flow-rate $q$ is seen as electric current. Typically, inlet flow-rate $q^i$ is measured (via magnetic resonance imaging or Doppler Ultrasound) and inlet pressure $P^i$ is measured by pressure catheters. The goal then is to estimate the parameters ($R_p$, $C$, and $R_d$) by assuming $q_i$ is deterministically known and minimising the difference between the $P^i$ reproduced by the model and the $P_i$ that was measured. The model dynamics is described by the following differential algebraic equations, which may also be rewritten as a single ordinary differential equation:
\begin{equation}
\begin{cases}
    \dot{P}^c & = \dot{P}_\mathrm{ext} + \dfrac{1}{C} (q^i - q^o) \\[5pt]
    P^c & = P_\mathrm{ven} + q^o \,R_d \\[5pt]
    P^i & = P^c + q^i \,R_p
\end{cases}, \qquad \text{ with } \qquad P^i(t=0) = P^i_0
    \label{eqn_RCR}
\end{equation}
where $P^i$ and $P^c$ are the inlet and mid-Windkessel pressures, respectively (Figure \ref{fig_RCR_schematic}); $P_\mathrm{ext}$ and $P_\mathrm{ven}$ are the reference external and venous pressures, respectively, which are both set to zero; and $q^i$ and $q^o$ are the inlet and outlet flow-rates, respectively.  The measurement model is written as follows:
\begin{equation}
    \label{eqn_RCR_obs}
    y_n = P^i_n + \epsilon_n \qquad \text{where}  \qquad \epsilon_n \sim \mathcal{N}(0,\sigma^2_\mathrm{noise})
\end{equation}

where $\epsilon_n$ is the  noise (normally distributed with zero mean and variance $\sigma_\mathrm{noise}^2$) in measuring $P_n^i$ to give the measurement $y_n$ at time $t_n$. The measurement vector, therefore, has only one component $\boldy_n = [y_n]$. The nominal values of $R_p$, $C$, $R_d$ are 0.838 mmHg$\cdot$s/cm$^3$, 0.0424 cm$^3$/mmHg, and 9.109 mmHg$\cdot$s/cm$^3$. Note that these units are chosen so that the results are comprehensible in typical units used in the clinic: ml for volume and mmHg for pressure. Figure \ref{fig_RCR_pq} shows the inlet flow-rate $q^i$ (taken from \cite{vignon2010outflow,pant2014methodological} where it was measured in the carotid artery of a healthy 27- year-old subject), and the resulting pressure curves obtained by the solution of equation \eqref{eqn_RCR} with $P^i_0 = 85$ mmHg and nominal parameter values. In order to put a zero-mean and  unit-variance prior on the parameters, see equation \eqref{eqn_prior_cov}, the following parameter transformation is considered
\begin{equation}
    \xi = \xi_0 + \varsigma_{\xi} \theta_{\xi} \qquad \text{ where } \qquad \xi \in \{R_p,C,R_d\}
\end{equation}

where $\xi$ represents the real parameter, $\xi_0$ and $\varsigma_\xi$ are transformation parameters, respectively, and $\theta_\xi$ represents the transformed parameter on which a prior of zero mean and unit variance is considered. Therefore, the prior considered on the real parameter $\xi$ has mean $\xi_0$ and variance $\varsigma_\xi^2$. The posterior variances for the transformed parameter $\theta_\xi$ and the real parameter $\xi$ are represented by $\sigma^2_\theta$ and $\sigma^2$, respectively.
A total of 150 time-points, evenly distributed between $t=0 $s and $t=T_c$ (where $T_c = 0.75 $s is the time period of the cardiac cycle), are used for the computation of information sensitivity functions and conditional variances.\\

Figure \ref{fig_RCR_marginal_var_isf} shows the marginal posterior variances (conditional only on the measurements, top row) and the corresponding information gains (bottom row) for individual parameters at four different levels of measurement noises. The conditional variances when all measurements are taken into account, i.e. at $t=T_c$, are also summarised in Table \ref{tab_RCR}. An immediate utility of Figure \ref{fig_RCR_marginal_var_isf} is in identify intervals of time where information is concentrated about a parameter. For example, from the first column it is clear that most of the information about the parameter $\theta_{R_p}$ is concentrated in the interval $t\in[0.3,0.4]$ as this is the interval that shows maximum reduction in the marginal variance and highest information gain. This interval corresponds to the rising peak of the inlet flow-rate curve, see Figure \ref{fig_RCR_pq}, and from equation \ref{eqn_RCR} it is clear that the parameter $\theta_{R_p}$ should have most effect on the pressure $P^i$ in this interval. For parameter $\theta_C$, it appears from Figure \ref{fig_RCR_marginal_var_isf} that while information is available in the entire cardiac cycle, larger amount of information is concentrated in the later half of the cardiac cycle, $t\in[0.4,0.75$. For $R_d$ information is available throughout the cardiac cycle. These observations have also been presented in \cite{pant2014methodological} through the computation of generalised sensitivity functions \cite{thomaseth1999generalized} and in \cite{pant2015Information} through a Monte Carlo type computation of information gain. However, as opposed to generalised sensitivity functions which can be non-monotonic and therefore hard to interpret, the information sensitivity functions are always monotonic. Furthermore, since the generalised sensitivity functions are normalised by design, they are forced to start at 0 and end at 1, thereby making the assessment of measurement noise difficult. On the other hand the effect of measurement noise is inherently built in to the information sensitivity functions.     
Figure \ref{fig_RCR_marginal_var_isf} quantifies how increasing measurement noise results in a decreasing amount of information gained about the parameters. While this behaviour is intuitively expected, its quantification with respect to each individual parameter is made possible with the proposed method. For example, while at $\sigma^2_\mathrm{noise} = 100.0$ mmHg$^2$ the conditional variance of the parameter $\theta_{R_p}$ after considering all the measurements is 0.158 square units, at $\sigma^2_\mathrm{noise} = 4900.0$ mmHg$^2$ this conditional variance is 0.887 square units. Comparing this to the prior variance of 1.0 square units, one may conclude that at measurement noise of 4900.0 mmHg$^2$ (standard deviation of 70.0 mmHg), the parameter $R_p$ is extremely difficult to identify relative to when the measurement noise is 100.0 mmHg$^2$ (standard deviation of 10.0 mmHg). A similar argument can be made for the parameter $\theta_C$, even though its identifiability is better than that of $\theta_{R_p}$ ($\theta_C$ has posterior variance of 0.672 square units at measurement noise of $\sigma^2_\mathrm{noise} = 4900.0$ mmHg$^2$). However, the parameter $\theta_{R_p}$ appears to be well identifiable even at $\sigma^2_\mathrm{noise} = 4900.0$ mmHg$^2$ with final posterior variance of 0.07 square units. This behaviour can be explained  by the fact that measurement noise is assumed to be independent and identically distributed with zero mean at all measurement times. Therefore, the mean pressure is measured much more precisely than individual pressure measurements, irrespective of the noise levels, as when mean/expectation of equation \eqref{eqn_RCR_obs} is taken, the expectation of noise component is zero:
\begin{equation}
\label{eqn_RCR_expectation}
    \mathbb{E}[y_n] = \mathbb{E} [P^i_n] + \mathbb{E}[\epsilon_n] = \mathbb{E} [P^i_n],
\end{equation}
where $\mathbb{E}$ denotes the expectation operator. From equation \eqref{eqn_RCR} and Figure \ref{fig_RCR_schematic}, the inlet mean pressure is equal to the inlet mean flow-rate times the sum of both resistances, \emph{i.e.} $\mathbb{E} [P^i_n] = (R_p + R_d)\; \mathbb{E}[q^i]$. Approximating $\mathbb{E}[y_n]$ by the sample mean as $(1/n)\sum_{0}^n y_i$, one obtains
\begin{equation}
    \label{eqn_RCR_sum}
    \frac{1}{n}\sum_{i=0}^n y_i = (R_p + R_d)\; \mathbb{E}[q^i].
\end{equation}

Since $q^i$ is assumed deterministic, from the above equation it can be seen that $R_p + R_d$ is indirectly measured with high precision. As $R_d$ is approximately an order of magnitude larger than $R_p$, it is natural that $R_d$ dominates the sum $(R_p + R_d)$ and hence, irrespective of the noise levels, a large amount of information is obtained about $R_d$ (Figure \ref{fig_RCR_marginal_var_isf}, last column). The order of magnitudes of the resistances are chosen by the physics of circulation, where the resistance of small vessels and microcirculation is significantly higher than that of large vessels \cite{vignon2010outflow,pant2014methodological}, and is reflected in the chosen priors for the problem.\\

\begin{figure}[htpb]
    \centering
    \includegraphics[width=1.0\linewidth]{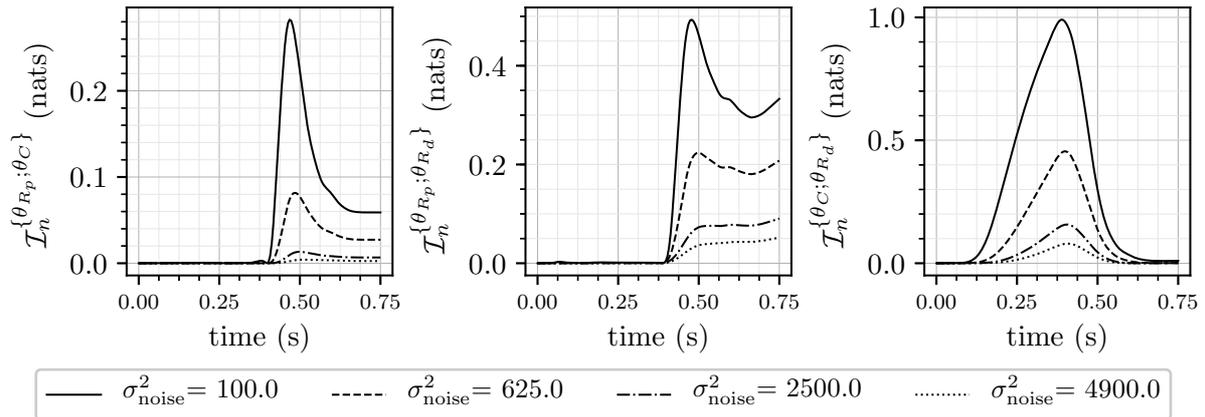}
    \caption{Mutual information between all the pairs of Windkessel model parameters at four different levels of measurement noise.}
    \label{fig_RCR_mi}
\end{figure}

Equation \eqref{eqn_RCR_sum} and the arguments presented above imply that a significant amount of correlation must have been built up between the parameters $R_p$ and $R_d$ in the posterior distribution as the sum $(R_p + R_d)$ is measured with high precision. This correlation implies that if one of the parameters $R_p$ or $R_p$ were known then how much additional information can be gained about the other parameter. The conditional mutual information (CMI) presented in equation \eqref{eqn_mi} precisely measures this additional information. CMIs for all the three pairs of the parameters are shown in Figure \ref{fig_RCR_mi}. It is clear that at the end of the cardiac cycle, the largest conditional mutual information is for the parameter pair $\theta_{R_p}$ and $\theta_{R_d}$. It is sensible to compare the magnitude of CMIs with the marginal information gains (Figure \ref{fig_RCR_marginal_var_isf}). For example, for the case of $\sigma^2_\mathrm{noise}=100.0$, the marginal gain in information about the parameter $R_p$ is approximately 0.9 nats and the mutual information between $R_p$ and $R_d$ is 0.35 nats; therefore, one may conclude that approximately 40\% extra information about the parameter $R_p$ is locked up in the correlation with $R_d$.  For the pair $R_d$ and $C$, it appears that correlation is built up in the $t\in[0.0,0.4]$, the diastole, and destroyed in the remaining part, the systole, of the cardiac cycle. This can be explained by the fact that the time-constant $e^{-t/\tau}$, with $\tau = R_d C$, is the dominant parameter that governs the diastole phase \cite{pant2015Information} leading to a built up of correlation, and as independent information about $C$ and $R_d$ is acquired in systole (Figure \ref{fig_RCR_marginal_var_isf}) this correlation is destroyed. It should be noted that these aspects, even without knowing the physics (or solution) of the problem, can be naturally inferred from Figures \ref{fig_RCR_marginal_var_isf} and \ref{fig_RCR_mi}.\\

\begin{figure}[htpb]
    \centering
    \includegraphics[width=1.0\linewidth]{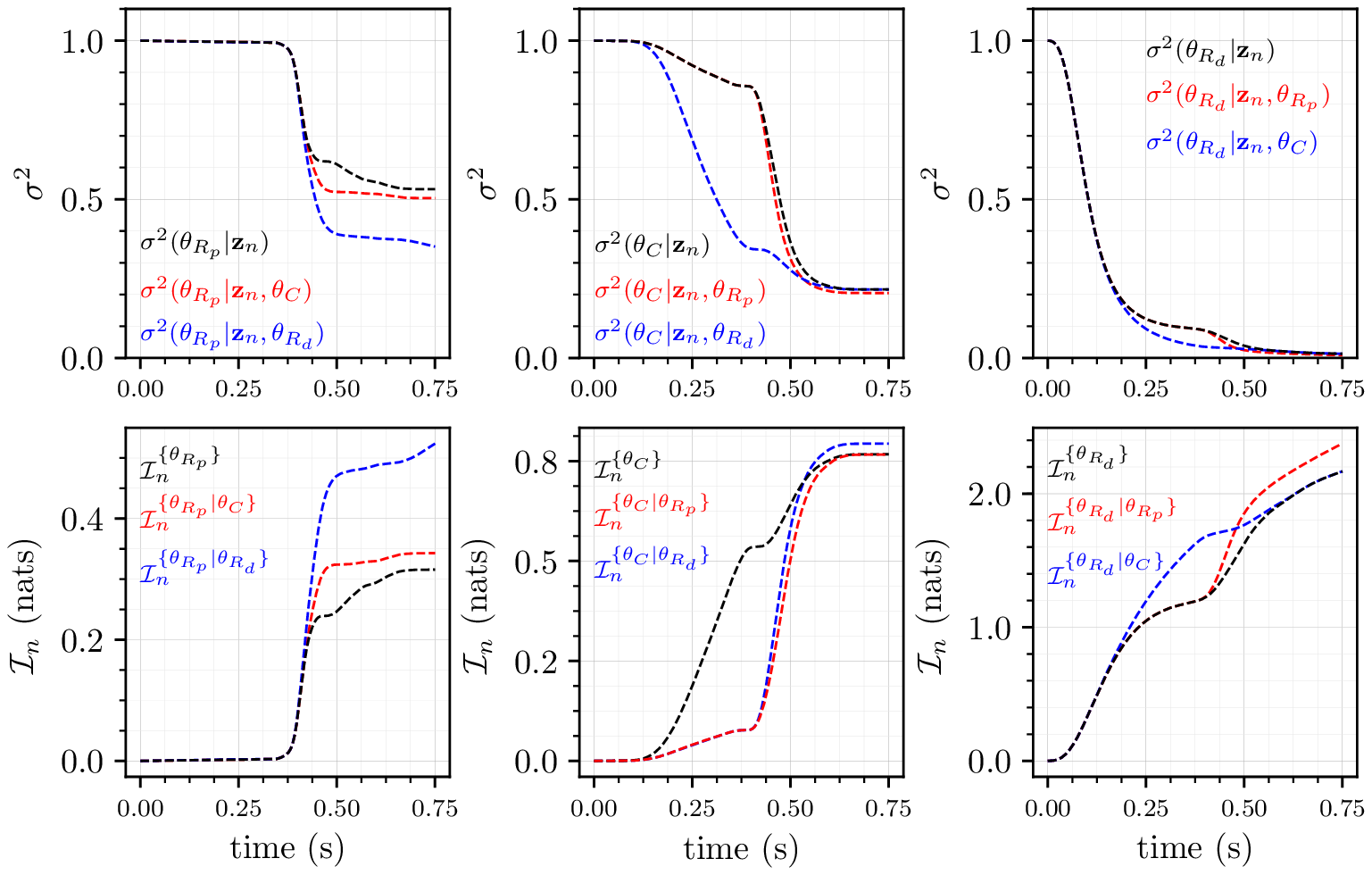}
    \caption{Conditional variances (top row) and conditional information gains (bottom row) for all pairs of the Windkessel model parameters. The measurement noise is $\sigma^2_{\mathrm{noise}}=625.0$.}
    \label{fig_RCR_conditional_var_isf}
\end{figure}

The effect of correlations can be further assessed by looking at the conditional variances (top row) and conditional information gains (bottom row) as depicted in Figure \ref{fig_RCR_conditional_var_isf}. For $\sigma^2_\mathrm{noise}=625.0$, this figure shows the conditional posterior variances and the conditional information gains for individual parameters when other parameters are given. For the parameter $\theta_{R_p}$, it can be seen that the conditional variance given $\theta_{R_d}$ is lower than the marginal variance in the interval $t\in[0.4,0.75]$ as this is the region where mutual information (correlation) is built between these parameters (Figure \ref{fig_RCR_mi}). Similarly, in diastole, $t\in[0.0,0.4]$, it can be seen that the conditional variance of parameter $\theta_C$ given $\theta_{R_d}$ is significantly lower as correlation is built up, but this gain quickly diminishes to zero in systole, $t\in[0.4,0.75]$. For the parameter $R_d$, as a large amount of individual information is obtained marginally, the conditional variances are not too different than the marginal variances. Note, that the variances show an opposite behaviour to information gains as a decrease in variance implies gain in information. Therefore, even though the two measures appear to be similar, information gain is a better measure as it can be readily applied to cases where behaviour of a set of parameters is required to be studied. For example if one was interested in the joint information again for a set of two parameters given a third, the information gain measure will be a scalar but the joint covariance will be a matrix.  Furthermore, the relation between conditional information gain, marginal information gain, and mutual information is additive, see equation \eqref{eqn_mi}, whereas the relation between conditional variance and marginal variance is, in general, not additive. As a demonstration, it can be observed that the conditional information gain curves in Figure \ref{fig_RCR_conditional_var_isf} can be obtained by the addition of the corresponding curves from Figures \ref{fig_RCR_marginal_var_isf} and \ref{fig_RCR_mi}.



\begin{table}[htbp]
  \centering
    \small
    \caption{Prior and posterior variances (marginal and conditional) for the Windkessel model }
    \begin{tabular}{|l|cccc|ccc|}
        \hline
        \textbf{Parameter} & \multicolumn{4}{c}{\textbf{Prior}}     & \multicolumn{3}{|c|}{\textbf{Expected posterior}} \\
          \hline
          & \multicolumn{2}{c|}{$\theta$-space} & \multicolumn{2}{c}{Real space ($\xi$)} & \multicolumn{1}{|l|}{$\theta$-space} & \multicolumn{2}{c|}{Real space ($\xi$)} \\
          \hline

          & \multicolumn{1}{l}{mean}       & \multicolumn{1}{l|}{variance}        & \multicolumn{1}{c}{mean}  & \multicolumn{1}{l|}{variance}   & \multicolumn{1}{l|}{variance }        & \multicolumn{1}{l}{variance}   & Std./prior-mean\\
          & \multicolumn{1}{c}{$\mu_\theta$} & \multicolumn{1}{c|}{$\sigma^2_\theta$} & \multicolumn{1}{c}{$\mu=\xi_0$} & \multicolumn{1}{c|}{$\sigma^2=\varsigma_\xi^2$} & \multicolumn{1}{c|}{$\sigma^2_\theta$} & \multicolumn{1} {c}{$\sigma^2$} & \multicolumn{1}{c|}{$\sigma/\xi_0$} \\[10pt]
        \hline
        \multicolumn{8}{|l|}{Observation noise, $\sigma^2_\mathrm{noise} = 100.0$} \\
        \hline
                   &     &     &          &          &          &          & \\
    $R_p$          & 0.0 & 1.0 & 8.40E-01 & 1.60E-01 & 1.58E-01 & 2.53E-02 & 18.9\% \\
    $R_p|C$        &     &     &          &          & 1.41E-01 & 2.25E-02 & 17.9\% \\
    $R_p|R_d$      &     &     &          &          & 8.13E-02 & 1.30E-02 & 13.6\% \\
                   &     &     &          &          &          &          & \\[-8pt]
    $C$            & 0.0 & 1.0 & 4.00E-02 & 4.00E-04 & 4.45E-02 & 1.78E-05 & 10.5\% \\
    $C|R_p$       &     &     &          &          & 3.95E-02 & 1.58E-05 & 9.9\% \\
    $C|R_d$       &     &     &          &          & 4.36E-02 & 1.74E-05 & 10.4\% \\
                   &     &     &          &          &          &          & \\[-8pt]
    $R_d$          & 0.0 & 1.0 & 9.11E+00 & 2.03E+01 & 2.73E-03 & 5.52E-02 & 2.6\% \\
    $R_d|R_p$      &     &     &          &          & 1.40E-03 & 2.84E-02 & 1.8\% \\
    $R_d|C$        &     &     &          &          & 2.67E-03 & 5.41E-02 & 2.6\% \\
                   &     &     &          &          &          &          & \\
        \hline
        \multicolumn{8}{|l|}{Observation noise, $\sigma^2_\mathrm{noise} = 625.0$} \\
        \hline
                  &     &     &          &          &          &          & \\
    $R_p$      & 0.0 & 1.0 & 8.40E-01 & 0.00E+00 & 5.32E-01 & 8.51E-02 & 34.7\% \\
    $R_p|C$    &     &     &          &          & 5.04E-01 & 8.06E-02 & 33.8\% \\
    $R_p|R_d$  &     &     &          &          & 3.51E-01 & 5.61E-02 & 28.2\% \\
               &     &     &          &          &          &          & \\[-8pt]
    $C$        & 0.0 & 1.0 & 4.00E-02 & 0.00E+00 & 2.16E-01 & 8.64E-05 & 23.2\% \\
    $C|R_p$   &     &     &          &          & 2.04E-01 & 8.18E-05 & 22.6\% \\
    $C|R_d$   &     &     &          &          & 2.16E-01 & 8.63E-05 & 23.2\% \\
               &     &     &          &          &          &          & \\[-8pt]
    $R_d$      & 0.0 & 1.0 & 9.11E+00 & 0.00E+00 & 1.31E-02 & 2.66E-01 & 5.7\% \\
    $R_d|R_p$  &     &     &          &          & 8.66E-03 & 1.75E-01 & 4.6\% \\
    $R_d|C$    &     &     &          &          & 1.31E-02 & 2.65E-01 & 5.7\% \\
                  &     &     &          &          &          &          & \\
        \hline
    \multicolumn{8}{|l|}{Observation noise, $\sigma^2_\mathrm{noise} = 2500.0$} \\
        \hline
                  &     &     &          &          &          &          & \\
    $R_p$       & 0.0 & 1.0 & 8.40E-01 & 0.00E+00 & 8.09E-01 & 1.29E-01 & 42.8\% \\
    $R_p|C$     &     &     &          &          & 7.98E-01 & 1.28E-01 & 42.5\% \\
    $R_p|R_d$   &     &     &          &          & 6.75E-01 & 1.08E-01 & 39.1\% \\
                &     &     &          &          &          &          & \\[-8pt]
    $C$         & 0.0 & 1.0 & 4.00E-02 & 0.00E+00 & 5.14E-01 & 2.05E-04 & 35.8\% \\
    $C|R_p$    &     &     &          &          & 5.07E-01 & 2.03E-04 & 35.6\% \\
    $C|R_d$    &     &     &          &          & 5.13E-01 & 2.05E-04 & 35.8\% \\
                &     &     &          &          &          &          & \\[-8pt]
    $R_d$       & 0.0 & 1.0 & 9.11E+00 & 0.00E+00 & 4.02E-02 & 8.15E-01 & 9.9\% \\
    $R_d|R_p$   &     &     &          &          & 3.36E-02 & 6.80E-01 & 9.0\% \\
    $R_d|C$     &     &     &          &          & 4.02E-02 & 8.13E-01 & 9.9\% \\
                  &     &     &          &          &          &          & \\
        \hline
        \multicolumn{8}{|l|}{Observation noise, $\sigma^2_\mathrm{noise} = 4900.0$} \\
        \hline
                  &     &     &          &          &          &          & \\
    $R_p$      & 0.0 & 1.0 & 8.40E-01 & 0.00E+00 & 8.87E-01 & 1.42E-01 & 44.8\% \\
    $R_p|C$    &     &     &          &          & 8.82E-01 & 1.41E-01 & 44.7\% \\
    $R_p|R_d$  &     &     &          &          & 7.99E-01 & 1.28E-01 & 42.6\% \\
               &     &     &          &          &          &          & \\[-8pt]
    $C$        & 0.0 & 1.0 & 4.00E-02 & 0.00E+00 & 6.72E-01 & 2.69E-04 & 41.0\% \\
    $C|R_p$   &     &     &          &          & 6.68E-01 & 2.67E-04 & 40.9\% \\
    $C|R_d$   &     &     &          &          & 6.70E-01 & 2.68E-04 & 40.9\% \\
               &     &     &          &          &          &          & \\[-8pt]
    $R_d$      & 0.0 & 1.0 & 9.11E+00 & 0.00E+00 & 7.05E-02 & 1.43E+00 & 13.1\% \\
    $R_d|R_p$  &     &     &          &          & 6.35E-02 & 1.29E+00 & 12.5\% \\
    $R_d|C$    &     &     &          &          & 7.03E-02 & 1.42E+00 & 13.1\% \\
        \hline
    \end{tabular}%
  \label{tab_RCR}%
\end{table}%

\subsection{The Hodgkin-Huxley model of a neuron}

The Hodgkin-Huxley model \cite{hodgkin1952quantitative} describes ionic exchanges and their relationship to the membrane voltage in a biological neuron. This model has also been used as the basis for several other ionic models to describe  a variety of excitable tissues such as cardiac cells \cite{noble1966applications}. The model is described by the following ordinary differential equations:
\begin{equation}
\begin{cases}
    \dot{V}_m & = \dfrac{1}{C_m} \left[ \iext - 
    \overbrace{\gna m^3 h(V_m - V_{Na})}^{\raisebox{3pt}{$I_{Na}$}}
    - \overbrace{\gk n^4 (V_m - V_K)} ^{\raisebox{3pt}{$I_{K}$}}
    - \overbrace{\gl (V_m - V_L)}^{\raisebox{3pt}{$I_{L}$}}
    \right] \\[5pt]
    \dot{m} & = \alpha_m(V_m) (1-m) - \beta_m(V_m) m \\[5pt]
    \dot{h} & = \alpha_h(V_m) (1-m) - \beta_h(V_m) h \\[5pt]
    \dot{n} & = \alpha_n(V_m) (1-m) - \beta_n(V_m) n \\[5pt]
\end{cases}
    \label{eqn_HH}
\end{equation}

with

\begin{equation}
\begin{cases}
    \alpha_m(V_m) = \dfrac{-0.1 (V_m+50)}{\exp\left(\frac{-(V_m + 50)}{10}\right)-1} \\[30pt]
     \beta_m(V_m) = 4 \exp\left(\dfrac{-(V_m+75)}{18}\right) \\[20pt]
    \alpha_h(V_m) = 0.07 \exp\left(\dfrac{-(V_m+75)}{20}\right) \\[20pt]
     \beta_h(V_m) = \dfrac{1.0}{\exp\left(\frac{-(V_m+45)}{10}\right)+1} \\[30pt]
    \alpha_n(V_m) = \dfrac{-0.01 (V_m+65)}{\exp\left(\frac{-(V_m+65)}{10}\right)-1} \\[30pt]
     \beta_n(V_m) = 0.125 \exp\left(\dfrac{-(V_m+75)}{80}\right)
\end{cases}
    \label{eqn_HH_alpha_beta}
\end{equation}

where $V_m$ is the membrane voltage, $C_m$ is the membrane capacitance, $I_\mathrm{ext}$ is the external current applied; $I_{Na}$, $I_K$, and $I_L$ are the sodium, potassium, and leakage currents, respectively; $V_{Na}$, $V_K$, and $V_L$ are the equilibrium potentials for sodium, potassium, and leakage ions, respectively;$g_{Na}$, $g_K$, and $g_L$ are the maximum conductances for the channels of sodium, potassium, and leakage ions, respectively; and $m$, $h$, and $n$ are the dimensionless gate variables, $m,h,n \in [0,1]$, that characterise the activation and inactivation of sodium and potassium channels. $C_m$ is set to 1 $\mu$F/cm$^2$, and the equilibrium potentials are defined  in milliVolts (mV) relative to the membrane resting potential, $E_R$, as follows \cite{lloyd2008cellml,hodgkincellml}:
\begin{equation}
\begin{cases}
    E_R      =  -75 \mathrm{\;mV} \\
    V_{Na}   =  E_R + 115 \mathrm{\;mV} \\
    V_K      =  E_R - 12 \mathrm{\;mV}\\
    V_L      =  E_R + 10.613 \mathrm{\;mV}
\end{cases}.
    \label{eqn_restingPotentials}
\end{equation}

\begin{figure}[htpb]
    \centering
    \includegraphics[width=1.0\linewidth]{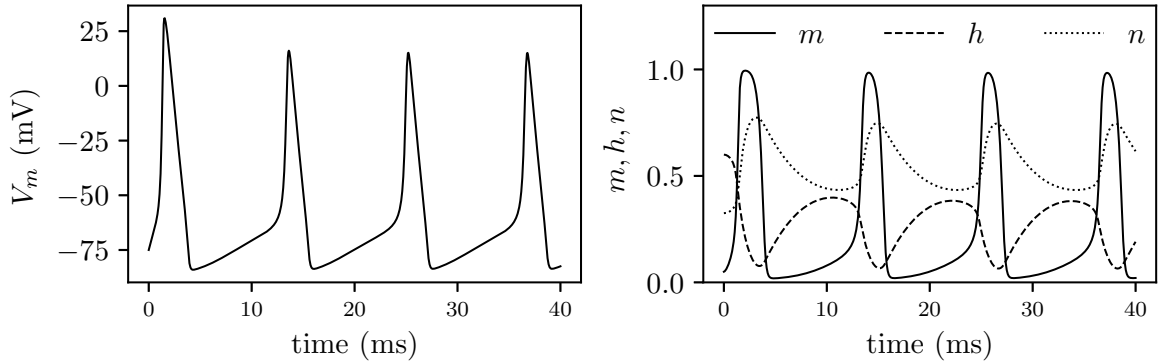}
    \caption{Solution of the Hodgkin-Huxley model, equation \eqref{eqn_HH}, for nominal parameter values.}
    \label{fig_HH_model}
\end{figure}

The inverse problem is of estimating the three parameters $g_{Na}$, $g_K$, and $g_L$ by measuring the membrane voltage $V_m$ when a constant external current $I_\mathrm{ext}=20 \mu$A/cm$^2$ is applied to the neuron. It is well known that when a relatively high constant external current is applied the neuron exhibits a tonic spiking pattern in membrane voltage $V_m$ \cite{izhikevich2004model,hasegawa2000responses,koch2004biophysics}. With nominal parameter values of $g_{Na}$=120.0 mS/cm$^2$, $g_{K}$=36.0 mS/cm$^2$, and $g_{L}$=0.3 mS/cm$^2$, and initial conditions of $V_m(0)=-75$ mV, $m(0)=0.05$, $h(0)=0.6$, and $n(0)=0.325$, this tonic spiking behaviour, generated by solving equation \eqref{eqn_HH}, is shown in Figure \ref{fig_HH_model}. The observation model reads
\begin{equation}
    \label{eqn_HH_obs}
    y_n = V_{m_n} + \epsilon_n \qquad \text{where}  \qquad \epsilon_n \sim \mathcal{N}(0,\sigma^2_\mathrm{noise})
\end{equation}
where $V_{m_n}$ is the membrane voltage at time $t_n$ and $\epsilon_n$ is the zero-mean measurement noise with variance $\sigma^2_\mathrm{noise}$. Since only $V_m$ is measured the observation vector is $\boldy_n=[y_n]$. As opposed to the Windkessel case where the effect of noise is evaluated, in this case the effect of number of observations, \emph{i.e.} the observation frequency is evaluated. $N_\mathrm{obs}$ number of measurement time-points evenly distributed in the time interval $t\in[0.0, 40.0]$ms are studied. Four levels of observation frequencies resulting in four values of $N_\mathrm{obs}\in\{100,200,400,800\}$ are used while $\sigma^2_\mathrm{noise}$ is set to 100.0 mV$^2$ (standard deviation of 10.0 mV). Similar to the Windkessel example the following parameterisation is used to impose zero-mean and unit-variance priors on the parameters.
\begin{equation}
    \xi = \xi_0 + \varsigma_{\xi} \theta_{\xi} \qquad \text{ where } \qquad \xi \in \{g_{Na},g_K,g_L\}
\end{equation}
where $\xi_0$ is the nominal parameter value, zero-mean and unit-variance normal distribution prior is imposed on the transformed parameter $\theta_\xi$, resulting in the prior disribution imposed on the real parameter $\xi$ to be a normal distribution with mean $\xi_0$ and variance $\varsigma^2_\xi$. The parameters $\varsigma_\xi$ are set to 10.0, 6.0, and 0.1 mS/cm$^2$ for $g_{Na}$, $g_K$, and $g_L$, respectively.\\

\begin{figure}[htpb]
    \centering
    \includegraphics[width=1.0\linewidth]{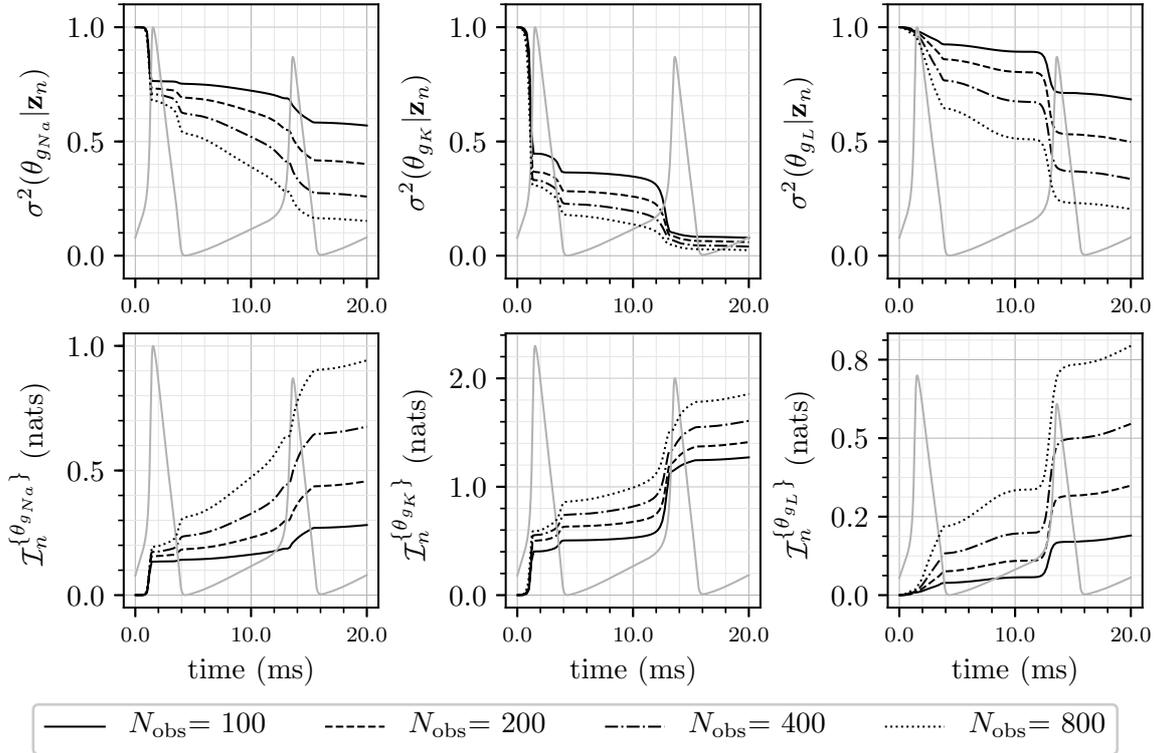}
    \caption{Marginal posterior variances (top row) and marginal information gains (bottom row) for all the Hodgkin-Huxley model parameters. Four different measurement frequencies are considered. In all the plots, an arbitrarily scaled $V_m$ curve is shown in grey.}
    \label{fig_HH_marginal_var_isf}
\end{figure}

\begin{figure}[tb]
    \centering
    \includegraphics[width=1.0\linewidth]{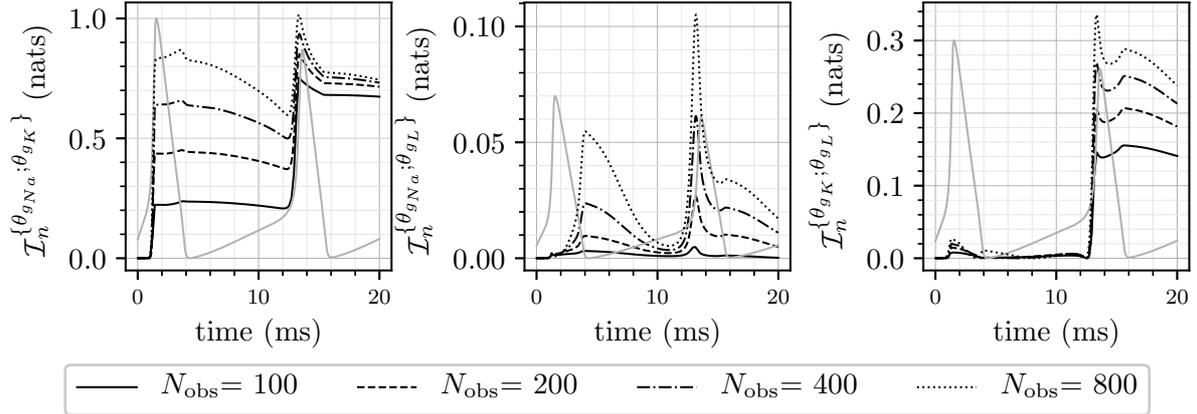}
    \caption{Mutual information between all the pairs of the parameters of the Hodgkin-Huxley model. Four different measurement frequencies are considered. In all the plots, an arbitrarily scaled $V_m$ curve is shown in grey.}
    \label{fig_HH_mi}
\end{figure}

Figure \ref{fig_HH_marginal_var_isf} shows the posterior marginal variances (top row) and the marginal information gains (bottom row) for the three parameters for all the four observation frequencies. In all these plots, an arbitrarily scaled $V_m(t)$ curve is shown in light grey for ease of interpretation relative to $V_m(t)$ variations. As expected, increasing the measurement frequency results in larger amounts of information (and consequently larger reduction in the posterior variances). However, it is observed that the parameters $\theta_{g_{Na}}$ and $\theta_{g_L}$ benefit most from an increase in measurement frequency as opposed to the parameter $\theta_{g_K}$ which benefits only marginally. This implies that at low observation frequencies the identifiability of $\theta_{g_K}$ is good, while very low amount of information is available for the parameters $\theta_{g_{Na}}$ and $\theta_{g_L}$. The behaviour for the parameter $\theta_{g_K}$ (middle column, Figure \ref{fig_HH_marginal_var_isf}) shows that the information about this parameter is concentrated mostly in the sharp rising phase of the action potential $V_m$. A similar behaviour, although less salient, is observed for the parameter $\theta_{g_{Na}}$ (first column, Figure \ref{fig_HH_marginal_var_isf}). While the Hodgkin-Huxley model is quite complex with gating variables of different time-constants and dependence of ionic currents on powers (up to fourth power) of the gating variables, it is widely understood that the rising phases of the action potential $V_m$ are related to the sodium and potassium currents. This may explain why information about the parameters $\theta_{g_{Na}}$ and $\theta_{g_K}$ is mostly concentrated in this region. Furthermore, if we accept that the sodium and potassium currents, in combination, are responsible for the rising action potential, then we should also expect a substantial amount of correlation between the parameters $\theta_{g_{Na}}$ and $\theta_{g_K}$ as it should be hard to distinguish between these two parameters. This is precisely what is observed by the conditional mutual information analysis, Figure \ref{fig_HH_mi}, where a large amount of mutual information is developed between these two parameters. For the case of $N_\mathrm{obs}=100$, the marginal information gain in the parameter $\theta_{g_{Na}}$, Figure \ref{fig_HH_marginal_var_isf}, is approximately 0.3 nats, and it is observed from Figure \ref{fig_HH_mi} that approximately 0.7 nats of mutual information exists between $\theta_{g_{Na}}$ and $\theta_{g_{K}}$. This implies that the amount of information that can be gained about $\theta_{g_{Na}}$ by knowing $\theta_{g_{K}}$, in addition to the measurements, is larger than the amount of information gained by just the measurements. Indeed, as the observation frequency is increased more information is available about all the parameters individually. Figure \ref{fig_HH_mi} also shows that significant amount of correlation is built between the parameters $\theta_{g_K}$ and $\theta_{g_L}$ during the sharp rising part of $V_m$. \\

\begin{figure}[tb]
    \centering
    \includegraphics[width=1.0\linewidth]{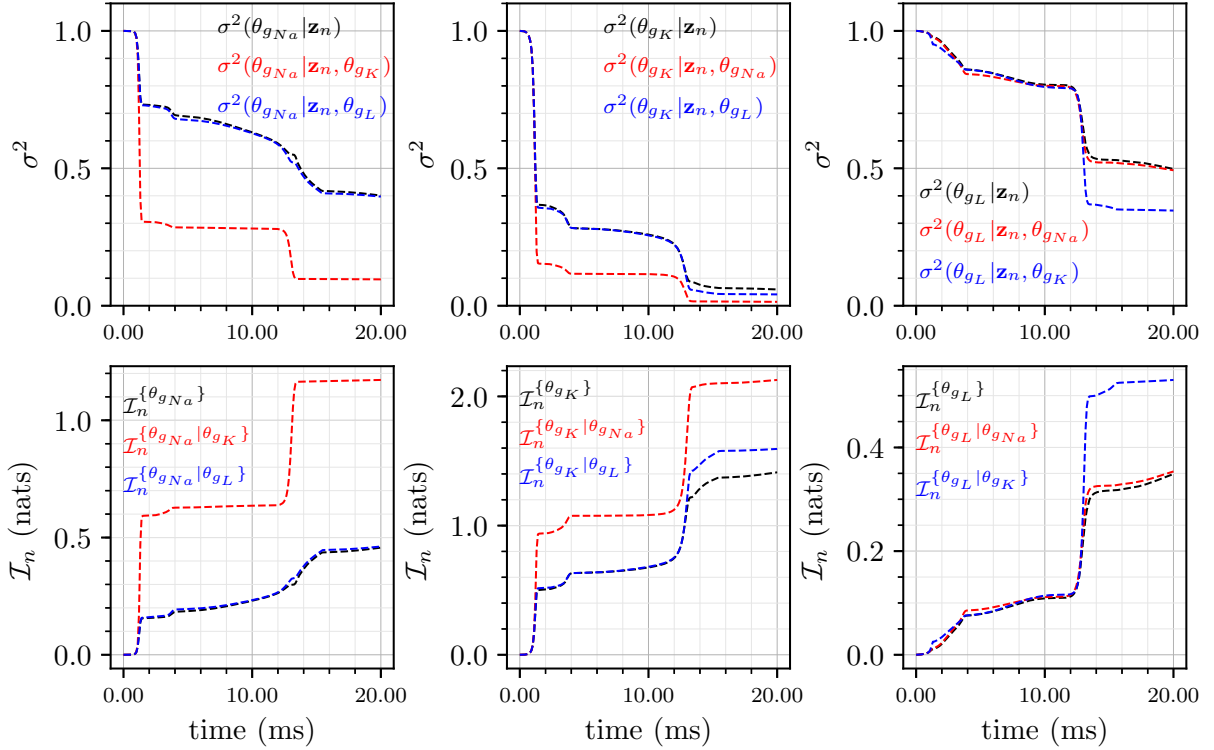}
    \caption{Conditional variances (top row) and conditional information gains (bottom row) for all pairs of the Hodgkin-Huxley model parameters. The case with $N_\mathrm{obs}=200$ is shown.}
    \label{fig_HH_conditional_var_isf}
\end{figure}

Finally, the effect of the conditional mutual information, \emph{i.e.} the correlation, can also be seen in terms of the conditional variances and conditional information gains as shown in Figure \ref{fig_HH_conditional_var_isf} for $N_\mathrm{obs}=200$. As discussed above the correlations between the pairs $(\theta_{g_{Na}}, \theta_{g_K})$ and $(\theta_{g_K},\theta_{g_L})$ show that the conditional variances are significantly lower (and the conditional information gain is larger) for one parameter when the other parameter is additionally known. It should be noted that the correlations and information gains presented are specific to the protocol, \emph{i.e} a constant external current resulting in tonic spiking of the neuron and only $V_m$ being measured. The information gains will behave differently if the protocol is changed, for example to a intermittent step currents or continuously varying external currents. Therefore, one application of the methods proposed in this article can be in optimal design of experiments, where one may design the protocol such that maximal information gain occurs for individual parameters while conditional mutual information (correlations in the posterior distribution) are minimised.  

\subsection{Influenza A virus kinetics}
\begin{figure}[htpb]
    \centering
    \includegraphics[width=1.0\linewidth]{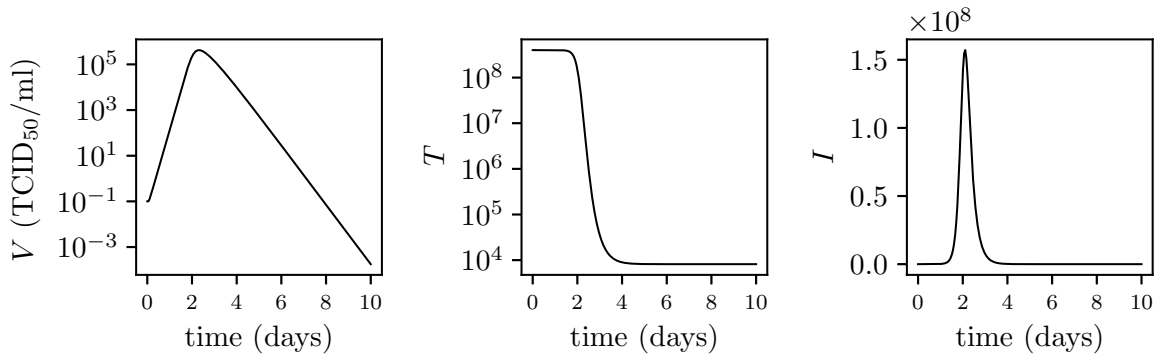}
    \caption{Solution of the Influenza A kinetics model, equation \eqref{eqn_influenza}, for nominal parameter values.}
    \label{fig_influenza_model}
\end{figure}

The final example presented is for the kinetics of the Influenza A virus. The following model was proposed by Baccam et. al. \cite{baccam2006kinetics} to describe viral infection

\begin{equation}
\begin{cases}
    \dot{V} & =  pI - cV \\[5pt]
    \dot{T} & = -\beta T V \\[5pt]
    \dot{I} & = \beta T V - \delta I
\end{cases}
    \label{eqn_influenza}
\end{equation}

where $V$ is the infectious virus titer (measured in TCID$_{50}$/ml of nasal wash), $T$ is the number of uninfected target cells, $I$ is the number of productively infected cells, and $\{\beta,\delta,p,c\}$ are the model parameters. The parameter $p$ represents the average rate at which the productively infected cells, $I$, increase the viral titers, and the parameter $\delta$ represents the rate at which the infected cells die. The parameter $\beta$ characterises the rate at which the susceptible cells become infected and $c$ represents the clearing rate of the virus.\\

As opposed to the previous example where the initial conditions were assumed to be known, in this example, the initial conditions for the virus titer $V_0$ and the number of uninfected target cells $T_0$ are considered unknown and hence form the parameters of the dynamical system. Time is measured in days (d) and the initial condition for the number of infected cells $I_0$ is assumed to be known at 0.0. Hence there are six parameters $[\beta,\delta,p,c,V_0,T_0]$ in total. The nominal values of the parameters are chosen to be $\beta = 2.7e-05$ (TCID$_{50}$/ml)$^{-1}$ d$^{-1}$, $\delta = 4.0$ d$^{-1}$, $p = 0.012$ TCID$_{50}$/ml$\cdot$d$^{-1}$, $c = 3.0$ d$^{-1}$, $V_0 = 0.1$ TCID$_{50}$/ml, and $T_0$ = 4e+08 based on the average patient parameters identified by Baccam et.~al. \cite{baccam2006kinetics}. As in the previous examples, the following parameterisation is used to impose zero-mean and unit-variance priors on the transformed parameters:

\begin{equation}
    \xi = \xi_0 + \varsigma_{\xi} \theta_{\xi} \qquad \text{ where } \qquad \xi \in \{\beta,\delta,p,c,V_0,T_0\}
\end{equation}

where $\theta_\xi$ represents the transformed version of the real parameter $\xi$, $\xi_0$ represents the nominal values of the parameter, and hence with a zero-mean and unit-variance prior on the transformed parameters $\theta_\xi$, the prior imposed on the real parameter is of mean $\xi_0$ and variance $\varsigma_\xi^2$. The scaling parameters $\varsigma_\xi$ are set to
9e-06,
1.3,
0.004,
1.0,
0.03, and
2.0e+08
for $\beta,\delta,p,c,V_0$, and $T_0$, respectively, in their respective units. The solution to equation \eqref{eqn_influenza} for the nominal parameter values is shown in Figure \ref{fig_influenza_model}. It is observed that both the virus titer $V$ and the number of infected cells $I$ increase sharply until they peak at the 2--3 day mark. After this a decrease in both values is observed. The number of uninfected target cells $T$ remains approximately constant until the 2 day mark after which a sharp decrease (approximately 4 orders of magnitude) is observed over the next two days leading to a plateau.

To study the sensitivity and information gain two cases are considered: first, when only $V$ is measured; and second, when both $V$ and $I$ are measured. In the first case, the observation model reads:
\begin{equation}
    \label{eqn_influenza_obs_V}
    y_n = V_n + \epsilon_n \qquad \text{where}  \qquad \epsilon_n \sim \mathcal{N}(0,\sigma^2_\mathrm{noise})
\end{equation}
where $V_n$ is the virus titer concentration at time $t_n$ and $\epsilon_n$ is the zero-mean measurement noise with variance $\sigma^2_\mathrm{noise} =$ 2.5e+07 (TCID$_{50}$/ml)$^2$, \emph{i.e.}, a standard deviation of 5e+03 TCID$_{50}$/ml. A total of 200 measurements are evenly distributed between 0 days and 10 days for the computation of marginal variances and information gains.

\begin{figure}[htpb]
    \centering
    \includegraphics[width=1.0\linewidth]{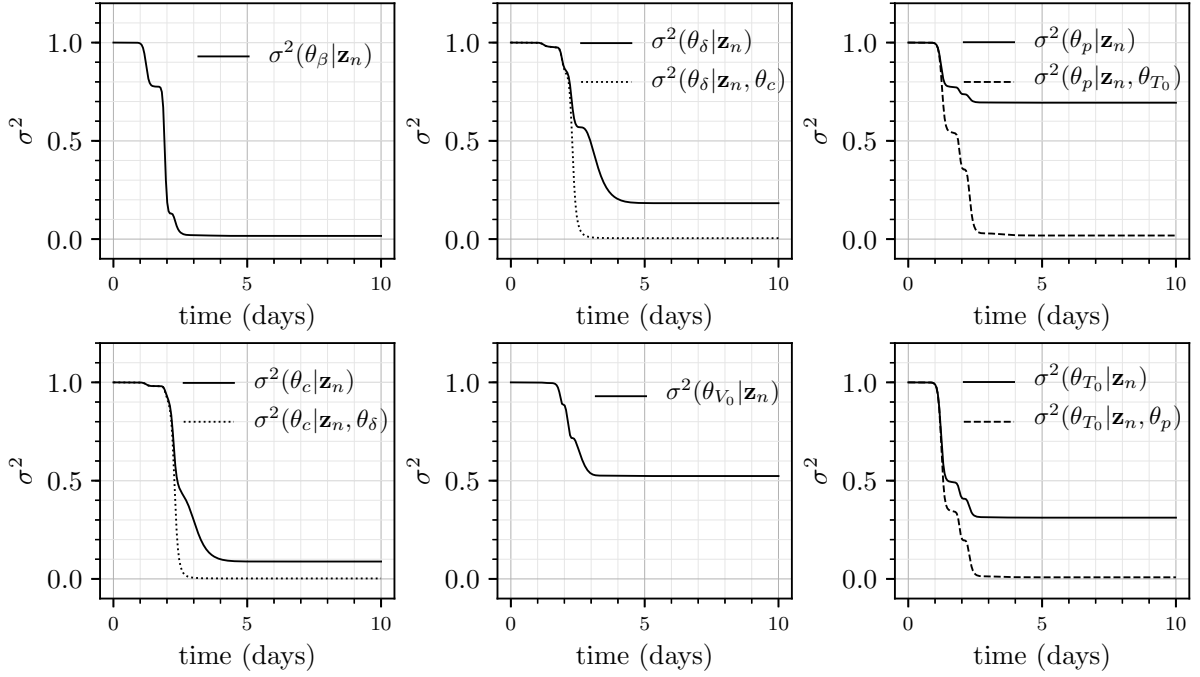}
    \caption{Marginal posterior variances (solid lines) and conditional variances (dashed lines) for the parameters of the Influenza A kinetics model. Only $V$ is measured with a measurement noise of $\sigma^2_\mathrm{noise} =$ 2.5e+07 (TCID$_{50}$/ml)$^2$.}
    \label{fig_influenza_var_1}
\end{figure}

\begin{figure}[htpb]
    \centering
    \includegraphics[width=1.0\linewidth]{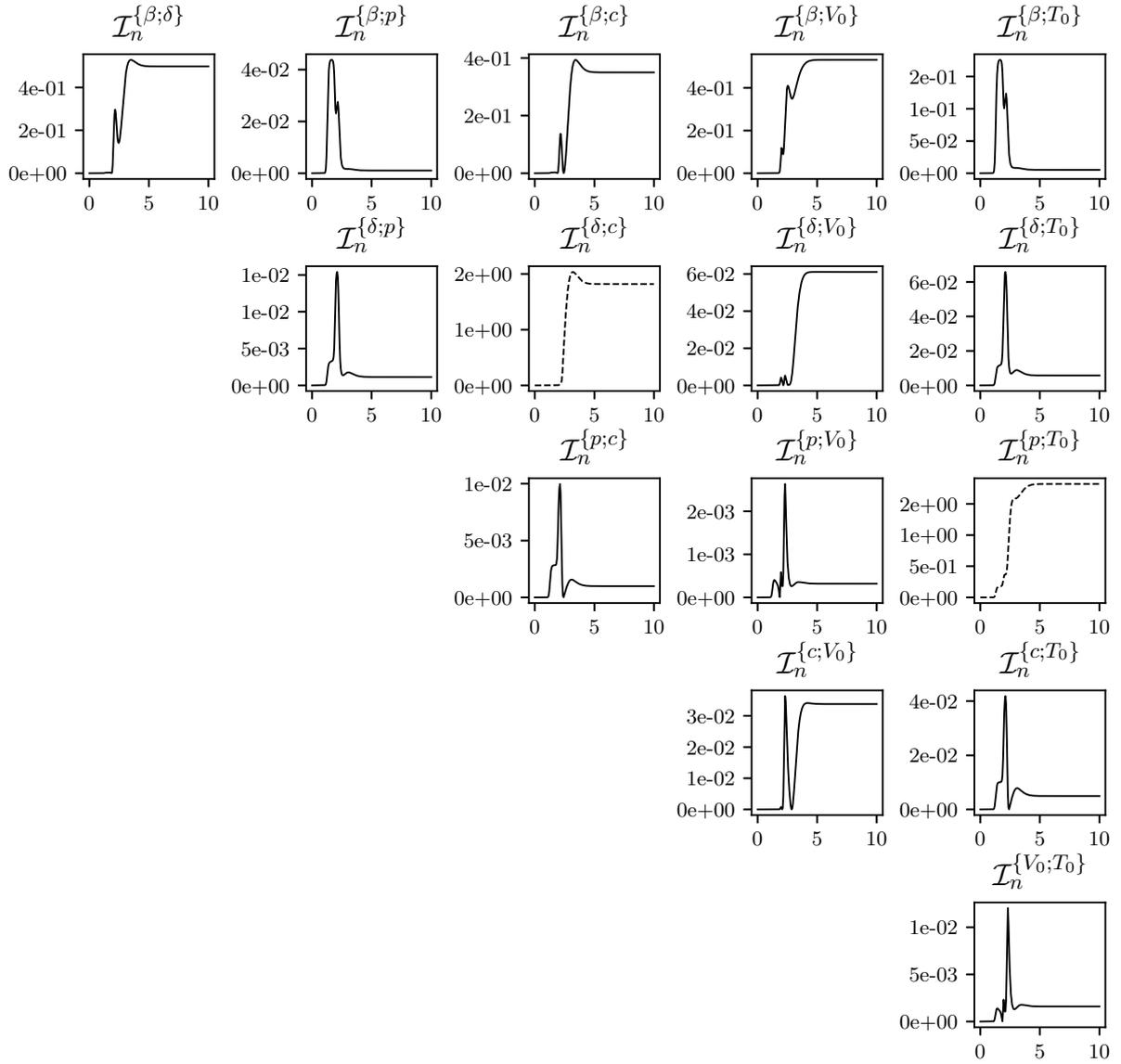}
    \caption{Mutual information between all pairs of the Influenza A kinetics model. Pairs with significant (large) mutual information are plotted in dashed lines.}
    \label{fig_influenza_mi}
\end{figure}

Figure \ref{fig_influenza_var_1} shows the marginal variances for all the parameters in solid lines and the conditional variances for a four pairs of parameters in dashed lines. Given the dynamics of the problem as shown in Figure \ref{fig_influenza_model} it is not surprising that most of the information gain about all the parameters occurs in $t\in[0,4]$ days. The parameters $\theta_\beta,\theta_\delta$, and $\theta_c$ appear to be well identifiable given the large decreases in marginal variances. However, the initial conditions $\theta_{V_0}$ and $\theta_{T_0}$ show less decrease in the variances indicating problems in their identifiability. Finally, the parameter $p$ appears to be unidentifiable given that its marginal variance decreases from 1.0 (standard deviation 1.0) to only 0.7 square units (standard deviation 0.84 units). Figure \ref{fig_influenza_mi} shows the mutual information between all the pairs of the parameters, where the parameter pairs that show a high mutual information are plotted in dashed lines. For the parameters in these pairs of high mutual information, $(\theta_\delta,\theta_c)$ and $(\theta_p,\theta_{T_0})$, the conditional variances are plotted in Figure \ref{fig_influenza_var_1}. The parameter pair $(\theta_p,\theta_{T_0})$ is particularly interesting as the parameter $\theta_p$, although unidentifiable individually, becomes very well identifiable, owing to the large mutual information it shares with $T_0$, if the initial condition $T_0$ is known. This observation was proved through classical methods by Miao et. al. \cite{miao2011identifiability} where it was shown that taking higher order derivatives of equation \eqref{eqn_influenza} and eliminating the unmeasured variables, $T$ and $I$, one obtains the following differential equation:
\begin{equation}
    \frac{d^3V}{dt^3} = 
    \left( \frac{d^2V}{dt^2}  + \delta c V + (\delta + c) \frac{dV}{dt} \right)
    \left( \frac{1}{V} \frac{dV}{dt} - \beta V \right)
    - \delta c \frac{dV}{dt}
    - (\delta + c)\frac{d^2V}{dt^2}.
    \label{eqn_influenza_third_derivative}
\end{equation}

\begin{figure}[htpb]
    \centering
    \includegraphics[width=0.9\linewidth]{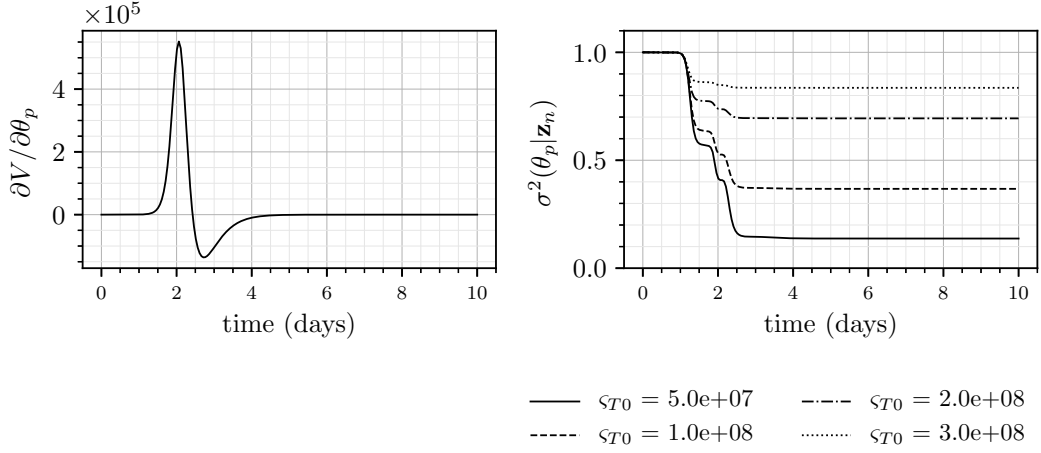}
    \caption{Sensitivity of the measurable $V$ with respect to $\theta_p$ (left panel) and the marginal posterior variance of the parameter $\theta_p$ at different levels of $\varsigma_{T_0}$ (right panel). The prior imposed on $T_0$ is of variance $\varsigma^2_{T_0}$.}
    \label{fig_influenza_var_2}
\end{figure}

\begin{figure}[htpb]
    \centering
    \includegraphics[width=1.0\linewidth]{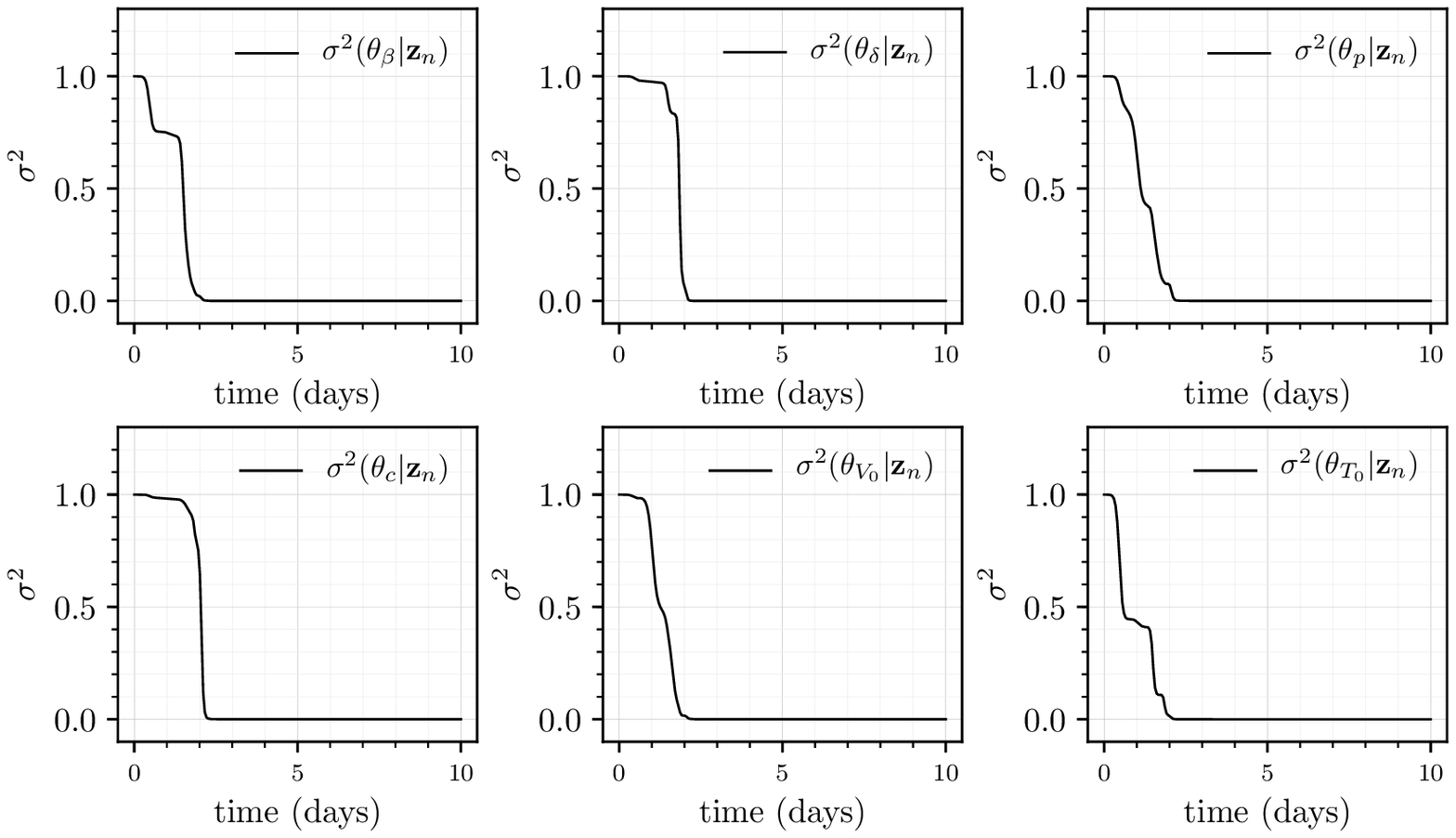}
    \caption{Marginal posterior variances for the parameters of the Influenza A kinetics model. when both $V$ and $I$ are measured.}
    \label{fig_influenza_var_3}
\end{figure}

Since the above equation does not contain the parameter $p$, in the absence of any other quantity, \emph{i.e.} $T$ and $I$, and the corresponding initial conditions, the parameter $p$ is not identifiable. Miao et.~al.~\cite{miao2011identifiability} also reported that when $T_0$ is known, the parameter $p$ becomes identifiable, which is consistent with the large mutual information. In the Bayesian approach adopted in this manuscript, a non-zero amount of knowledge (non-infinite variance) is inherently assumed in the prior for $\theta_{T_0}$, which results in a small amount of information gain (and hence a small reduction in the marginal variance from 1.0 to 0.7 square units). This small amount of information gain is a result of the knowledge assumed in the prior. However, it is not significant enough to hide the identifiability problem for $\theta_{p}$. One can choose to impose prior of higher ignorance by increasing the prior variance of the real parameter ${T_0}$ by increasing the scaling factor $\varsigma_{T_0}$. The results for four different values of $\varsigma_{T_0}$ on the marginal variance of the parameter $\theta_p$ are shown in Figure \ref{fig_influenza_var_2} (right panel). It is clear that a higher value of $\varsigma_{T_0}$, which implies higher ignorance in the prior for $T_0$, results in a decreasing amount of information gained about the parameter $\theta_p$. This example shows how, without the use of classical analytical methods, see for example those presented in \cite{miao2011identifiability}, which may not be easily applicable to all dynamical systems, the information theoretic approach can provide similar conclusions about parameter identifiability. Lastly, the classical sensitivity of the parameter $p$ to the measurable $V$ is shown in Figure \ref{fig_influenza_var_2} (left panel), whose large magnitude does not indicate any problems of parameter identifiability. Finally, Miao et.~al.~\cite{miao2011identifiability} reported that all the parameters of the influenza dynamical system were well identifiable if both $V$ and $I$, or both $V$ and $T$ were measured. For the case when both $V$ and $I$ are measured, the marginal variances are shown in Figure \ref{fig_influenza_var_3}, which too shows that no identifiability problems persist in this case. Note that the error structure in the measurement of $I$ was assumed to be identical to the measurement of $V$, equation \eqref{eqn_influenza_obs_V}.

\section{Conclusions}
A new class of functions called the `information sensitivity functions' have been proposed to study parametric information gain in a dynamical system. Based on a Bayesian and information-theoretic approach, such functions are easy to compute through classical sensitivity analysis. Compared to the previously proposed generalized sensitivity functions (GSFs) \cite{thomaseth1999generalized} to measure such information gain, the ISFs do not suffer from the forced-to-one behaviour and are easy to interpret as correlations are measured through separate measures of mutual information as opposed to oscillations in GSFs. Furthermore, as opposed to GSFs, which are normalised, the ISFs can be used to compare information gain between different parameters and hence can be used to rank the parameters on ease of identifiability. They can be used to identify regions of high information content and indicate identifiability problems for parameters which show little to no information gain, or high mutual information (correlation) with other parameters. The application of ISFs is demonstrated on three models. For the Windkessel model, the effect of measurement noise is illustrated and it is shown that the insights provided by ISFs are consistent with those of a significantly more expensive Monte Carlo type approach \cite{pant2015Information}. For the Hodgkin-Huxley model the effect of measurement frequency is illustrated, and finally, for the Influenza A virus, it is shown how, even when classical sensitivity analysis fails to assess identifiability issues, the ISFs correctly reveal identifiability problems, which have been analytically proven through classical methods.

\section*{Acknowledgements}
The author would like to thank Dr.~Damiano Lombardi (Inria Paris) for fruitful discussions and insights.


\appendix
\section{Differential analysis for equivalence of sensitivity and covariance evolution}
\label{app_differential_analysis}

From equation \eqref{eqn_discretised_forward_mean_app}, the linearised dynamical system is
\begin{equation}
    \dot{\mathbf{x}} = \mathbf{f}\Bigr\rvert_{n}  
    + \nabla_{\mathbf{x}} \mathbf{f}\Bigr\rvert_{n} \left( \mathbf{x} - \boldsymbol{\mu}_{\mathbf{x}_{n}}  \right)   
    + \nabla_{\boldsymbol{\theta}} \mathbf{f}\Bigr\rvert_{n} \left(\boldsymbol{\theta} - \boldsymbol{\mu}_{\boldsymbol{\theta}} \right)
    + \gradft\Bigr\rvert_{n} (t-t_n) 
\end{equation}
Separating the random variables $\btheta$ and $\boldx$ gives
\begin{equation}
    \dot{\mathbf{x}} = 
    \nabla_{\mathbf{x}} \mathbf{f}\Bigr\rvert_{n} \boldx
    + \nabla_{\boldsymbol{\theta}} \mathbf{f}\Bigr\rvert_{n} \boldsymbol{\theta}
    + \underbrace{\left(  \mathbf{f}\Bigr\rvert_{n}
    - \nabla_{\mathbf{x}} \mathbf{f}\Bigr\rvert_{n} \boldsymbol{\mu}_{\mathbf{x}_{n}}
    - \nabla_{\boldsymbol{\theta}} \mathbf{f}\Bigr\rvert_{n} \boldsymbol{\mu}_{\boldsymbol{\theta}}
    + \gradft\Bigr\rvert_{n} (t-t_n) \right)}_{\boldsymbol{\tau}_n}.
\end{equation}

Combined with trivial dynamics for the parameters $\dot{\btheta} = 0$, the dynamics for the combined vector $[\boldx^{\T}, \btheta^{\T}]^{\T}$ can be written as
\begin{equation}
    \underbrace{\begin{bmatrix}
        \dot{\boldx}\\[1ex]
            \dot{\btheta}
    \end{bmatrix}}_{\dot{\pmb{\boldsymbol{\varkappa}}}}
    =
    \underbrace{\begin{bmatrix}
        \nabla_{\mathbf{x}} \mathbf{f}\Bigr\rvert_{n} & \nabla_{\boldsymbol{\theta}} \mathbf{f}\Bigr\rvert_{n} \\[1ex]
          \zero_{p,d} & \zero_{p,p}
    \end{bmatrix}}_{\boldsymbol{\mathcal{F}}_n}
        \underbrace{\begin{bmatrix}
        {\boldx}\\[1ex]
            {\btheta}
        \end{bmatrix}}_{\pmb{\boldsymbol{\varkappa}}}
    + 
    \underbrace{\begin{bmatrix}
        {\boldsymbol{\tau}_n}\\[1ex]
            \zero_{p,1}
    \end{bmatrix}}_{\boldsymbol{\mathfrak{r}_n}}
\end{equation}
where $\zero_{a,b}$ represents a zero matrix of size $a\times b$. The above is concisely written as
\begin{equation}
    \dot{\pmb{\boldsymbol{\varkappa}}} = \boldsymbol{\mathcal{F}}_n \pmb{\boldsymbol{\varkappa}} + \boldsymbol{\mathfrak{r}_n}
\end{equation}
For the above stochastic differential equation, it is well known, see for example \cite{gelb1974applied}, that the covariance matrix of $\bkp$, denoted by $\bxi$, evolves according to the following differential equation 
\begin{equation}
    \label{eqn_cov_ode}
    \dot{\bxi} = \boldsymbol{\mathcal{F}}_n \; \bxi + \bxi \; \boldsymbol{\mathcal{F}}_n^{\T}.
\end{equation}
Therefore, if the covariance matrix of $\bkp_n$ is 
\begin{equation}
   \bxi = \cov \left(
    {\begin{bmatrix}
        {\boldx}\\[1ex]
            {\btheta}
        \end{bmatrix}}
        \right)
        =
\begin{bmatrix}
    \boldsymbol{\Sigma}_{n,n} & \boldsymbol{\Lambda}_{n,\btheta} \\[1ex]
           \boldsymbol{\Lambda}_{\btheta,n}^{\T} & \boldsymbol{\Sigma}_{\btheta,\btheta}
        \end{bmatrix},
\end{equation}
then $\bxi$ evolves according to equation \eqref{eqn_cov_ode} as
\begin{align}
    \dot{\bxi} & =
\begin{bmatrix}
        \nabla_{\mathbf{x}} \mathbf{f}\Bigr\rvert_{n} & \nabla_{\boldsymbol{\theta}} \mathbf{f}\Bigr\rvert_{n} \\[1ex]
          \zero_{p,d} & \zero_{p,p}
        \end{bmatrix}
\begin{bmatrix}
    \boldsymbol{\Sigma}_{n,n} & \boldsymbol{\Lambda}_{n,\btheta} \\[1ex]
           \boldsymbol{\Lambda}_{\btheta,n} & \boldsymbol{\Sigma}_{\btheta,\btheta}
        \end{bmatrix}
        +
\begin{bmatrix}
    \boldsymbol{\Sigma}_{n,n} &\boldsymbol{\Lambda}_{n,\btheta} \\[1ex]
           \boldsymbol{\Lambda}_{\btheta,n} & \boldsymbol{\Sigma}_{\btheta,\btheta}
        \end{bmatrix}
\begin{bmatrix}
    \nabla_{\mathbf{x}}^{\T} \mathbf{f}\Bigr\rvert_{n} & \zero_{d,p}  \\[1ex]
          \nabla_{\boldsymbol{\theta}}^{\T} \mathbf{f}\Bigr\rvert_{n} & \zero_{p,p}
\end{bmatrix} \nonumber \\[1ex]
     &= 
\begin{bmatrix}
    \nabla_{\mathbf{x}} \mathbf{f}\Bigr\rvert_{n} \boldsymbol{\Sigma}_{n,n} + \nabla_{\boldsymbol{\theta}} \mathbf{f}\Bigr\rvert_{n} \boldsymbol{\Lambda}_{n,\btheta}^{\T} + \boldsymbol{\Sigma}_{n,n} \nabla_{\mathbf{x}}^{\T} \mathbf{f}\Bigr\rvert_{n} + \boldsymbol{\Lambda}_{n,\btheta} \nabla_{\boldsymbol{\theta}}^{\T} \mathbf{f}\Bigr\rvert_{n}
        & \phantom{x}\nabla_{\mathbf{x}} \mathbf{f}\Bigr\rvert_{n} \boldsymbol{\Lambda}_{n,\btheta}  + \nabla_{\boldsymbol{\theta}} \mathbf{f}\Bigr\rvert_{n} \boldsymbol{\Sigma}_{\btheta,\btheta} \\[2ex]
           \boldsymbol{\Lambda}_{n,\btheta}^{\T} \nabla_{\mathbf{x}}^{\T} \mathbf{f}\Bigr\rvert_{n} + \boldsymbol{\Sigma}_{\btheta,\btheta} \nabla_{\boldsymbol{\theta}}^{\T} \mathbf{f}\Bigr\rvert_{n} & \zero_{p,p}
\end{bmatrix}. \label{eqn_cov_evolution_expanded}
\end{align}
The next task is to relate the above evolution of the covariance matrix with the evolution of sensitivity matrix. From equation \eqref{eqn_sensitivity} the sensitivity matrix $\boldS$ evolves as
\begin{equation}
    \label{eqn_sens_1}
    \dot{\boldS} = \nabla_{\mathbf{x}} \mathbf{f}\Bigr\rvert_{n} \mathbf{S}_{n} + \nabla_{\boldsymbol{\theta}} \mathbf{f}\Bigr\rvert_{n}   
\end{equation}
Therefore, taking the transpose of equation \eqref{eqn_sens_1} yields
\begin{equation}
    \label{eqn_sens_2}
    (\dot{\boldS})^{\T} = \left(\frac{d \boldS}{dt}\right)^{\T}
    = \frac{d (\boldS^{\T})}{dt}
    = \mathbf{S}_{n}^{\T} \nabla_{\mathbf{x}}^{\T} \mathbf{f}\Bigr\rvert_{n} + \nabla_{\boldsymbol{\theta}}^{\T} \mathbf{f}\Bigr\rvert_{n}.
\end{equation}
The derivative of the matrix product $\boldS \boldS^{\T}$ can be written as follows
\begin{equation}
    \frac{d (\boldS \boldS^{\T})}{dt} = \boldS \frac{d (\boldS^{\T})}{dt} + \frac{d (\boldS)}{dt} \boldS^{T}
\end{equation}
Substituting the derivatives from \eqref{eqn_sens_1} with \eqref{eqn_sens_2} into the above equation gives
\begin{equation}
    \frac{d (\boldS \boldS^{\T})}{dt} =   \boldS \boldS^{\T} \nabla_{\mathbf{x}}^{\T} \mathbf{f}\Bigr\rvert_{n} + \nabla_{\mathbf{x}} \mathbf{f}\Bigr\rvert_{n} \boldS \boldS^{\T} + \nabla_{\boldsymbol{\theta}} \mathbf{f}\Bigr\rvert_{n} \boldS^{\T} + \boldS \nabla_{\boldsymbol{\theta}}^{\T} \mathbf{f}\Bigr\rvert_{n}.
\end{equation}
It is easy to see that if $\boldsymbol{\Sigma}_{\btheta,\btheta} = \eye_p$, $\boldsymbol{\Lambda}_{n,\btheta} = \boldS$, and $\boldsymbol{\Sigma}_{n,n} = \boldS \boldS^{\T}$, then the state covariance, $\boldsymbol{\Sigma}_{n,n}$, and the cross-covariance, $\boldsymbol{\Lambda}_{n,\btheta}$, from equation \eqref{eqn_cov_evolution_expanded} evolve as
\begin{equation}
    \dot{\boldsymbol{\Sigma}}_{n,n} = \boldS \boldS^{\T} \nabla_{\mathbf{x}}^{\T} \mathbf{f}\Bigr\rvert_{n} + \nabla_{\mathbf{x}} \mathbf{f}\Bigr\rvert_{n} \boldS \boldS^{\T} + \nabla_{\boldsymbol{\theta}} \mathbf{f}\Bigr\rvert_{n} \boldS^{\T} + \boldS \nabla_{\boldsymbol{\theta}}^{\T} \mathbf{f}\Bigr\rvert_{n} = \frac{d (\boldS \boldS^{\T})}{dt},
\end{equation}
and
\begin{equation}
    \dot{\boldsymbol{\Lambda}}_{n,\btheta} =  \nabla_{\mathbf{x}} \mathbf{f}\Bigr\rvert_{n} \mathbf{S}_{n} + \nabla_{\boldsymbol{\theta}} \mathbf{f}\Bigr\rvert_{n} = \dot{\boldS}.
\end{equation}
\end{document}